\documentclass[11pt]{article}

\usepackage[
    a4paper,
    top=2.0cm,
    bottom=2.0cm,
    left=2.15cm,
    right=2.15cm
]{geometry}

\usepackage{amsmath,amssymb,amsthm,mathtools}
\usepackage{setspace}
\usepackage{microtype}
\usepackage{needspace}
\usepackage[authoryear,round]{natbib}
\usepackage[hidelinks]{hyperref}
\usepackage[nameinlink,capitalize,noabbrev]{cleveref}
\usepackage{lmodern}
\usepackage[T1]{fontenc}
\usepackage{silence}
\crefname{appendix}{appendix}{appendices}
\Crefname{appendix}{Appendix}{Appendices}

\makeatletter
\numberwithin{equation}{section}
\theoremstyle{plain}
\newtheorem{lem}{\protect\lemmaname}
\theoremstyle{plain}

\theoremstyle{plain}
\newtheorem{thm}{\protect\theoremname}
\theoremstyle{plain}
\newtheorem{prop}{Proposition}
\theoremstyle{definition}

\bibpunct{[}{]}{;}{y}{}{,}

\makeatother

\usepackage{babel}
\providecommand{\corollaryname}{Corollary}
\providecommand{\lemmaname}{Lemma}
\providecommand{\theoremname}{Theorem}

\begin{document}

\title{\textbf{Auction Design with Private Priors and Costly Learning}}
\author{Kemal Ozbek\thanks{Department of Economics, University of Southampton, University Road,
Southampton, SO17 1BJ, United Kingdom. Email:mkemalozbek@gmail.com}}
\maketitle

\begin{abstract}
We study two-stage auctions in which bidders privately know their valuation priors and learning costs, while the seller jointly designs information acquisition, continuation allocation, and payments. Revenue maximization requires screening learning costs while satisfying cross-prior incentives, potentially distorting both experimentation and continuation allocation. We derive a revenue representation involving allocated posterior values, virtual learning costs, and prior-dependent rents, and characterize cross-prior incentives through difference inequalities. We establish existence of optimal mechanisms for finite experiment menus and arbitrary prior distributions. With finite prior support, bounded finite-category linear programs approximate optimal revenue at a linear rate while preserving exact dynamic incentive compatibility and interim individual rationality. Their dual yields joint optimality conditions for experimentation, allocation, and payments. A two-prior example demonstrates that efficient continuation can be revenue-suboptimal and that cost-contingent experimentation can strictly increase revenue.
\end{abstract}



\section{Introduction}

A bidder may enter an auction uncertain about the value
of the object, while privately knowing both her prior
beliefs and the cost of learning more. In a corporate
acquisition, for example, a potential buyer may privately
assess the target's prospects and the cost of conducting
due diligence. The seller may control access to data rooms,
management presentations, and specialist reports before
final bids are submitted.\footnote{
Competitive acquisition processes frequently involve
staged access to information and management.
For example, during the 2007 sale of Manor Care,
prospective acquirers participated in management
presentations and extensive due diligence before
submitting final acquisition proposals
\citep[Appendix~A]{GorbenkoMalenko2014}.
Similarly, during the 2024 sale of AssetMark
Financial Holdings, five bidders selected for
the second round received access to a detailed
virtual data room, and the subsequent bidding
process involved management meetings and
due-diligence sessions
\citep[p.~26]{AssetMark2024}.
Both transactions illustrate how sellers can coordinate
bidders' opportunities to acquire information before
final offers are submitted.
}
The seller therefore faces a joint design problem:
which information should bidders acquire, and how
should the object and payments be allocated after
that information is obtained?

We study this problem in a two-stage auction of a
single indivisible object. Each bidder has a binary
valuation $V_i\in\{a,b\}$ and privately observes
a prior $s_i=\Pr(V_i=b)$ and a learning-cost scale
$r_i$. An assigned experiment $\pi_i$ costs
$r_iC(\pi_i)$. In the first stage, the seller elicits
private types and assigns experiments, potentially
coordinating learning across bidders. Following
experimentation, a continuation mechanism determines
allocation and payments. Both stages are chosen
to maximize expected seller revenue.

The central difficulty is that screening learning
costs cannot be separated from eliciting private
priors. A bidder's prior determines the expected
benefit of information and her incentives to imitate
other prior types. Such imitation may change not only
the experiment assigned to her but also the
continuation allocation and payment rules she faces.
Consequently, the seller must jointly control
within-prior cost incentives and cross-prior
deviations. In particular, efficient continuation
allocation need not maximize revenue: distorting
allocation after learning may reduce the rents
required to induce truthful prior reports.
We therefore allow general report-contingent
continuation allocation and signed payments rather
than imposing VCG or efficient continuation.

Our first contribution is a revenue representation
that separates the value of allocated information,
virtual learning costs, and prior-dependent rents:
\[
\operatorname{Rev}(M)
=
E\left[
\sum_i p_i^M T_i
-
\sum_i\psi(r_i)C(\pi_i^M)
\right]
-
\sum_i E[K_i^M(s_i)],
\]
where $T_i$ is bidder $i$'s posterior expected
valuation, $p_i^M$ is her allocation probability,
$K_i^M(s_i)$ is her boundary rent, and
\(
\psi(r)=r+F_r(r)/f_r(r)
\)
is the virtual learning-cost scale.
Thus welfare evaluates experimentation at actual
costs, whereas revenue evaluates it at virtual
costs. Unlike a standard one-dimensional screening
problem, however, boundary rents across prior types
are linked by incentive constraints. With finite
prior support, we characterize these constraints
through directed difference inequalities, providing
a tractable description of the additional restrictions
imposed by private priors.

Our second contribution concerns the existence and
construction of revenue-maximizing mechanisms.
Within the maintained mechanism class, bounded
continuation payments suffice without revenue loss,
and optimal mechanisms exist for finite experiment
menus and arbitrary prior distributions.
For finite prior and experiment sets, we construct
finite-category linear programs whose optimal values
converge to optimal revenue at a rate proportional
to the maximum cost-cell length. Importantly, the
corresponding mechanisms satisfy exact dynamic
incentive compatibility and interim individual
rationality for every cost type, rather than only
at the discretization points. The dual provides
joint optimality conditions for experiment
assignments, continuation allocations, and payments,
revealing how incentive constraints affect the
seller's choices at both stages.

A two-prior example illustrates the distinction
between efficiency and revenue maximization.
Even with fixed experimentation, efficient
continuation allocation can be strictly
revenue-suboptimal. Moreover, allowing experiments
to depend on reported priors alone yields no revenue
improvement over optimal fixed experimentation in
the example, whereas conditioning them on reported
learning costs produces a strict gain. The example
thus shows how selective information acquisition
and distorted continuation allocation can work
together to improve revenue.

Our analysis contributes to the literature on
auctions with costly learning and seller-controlled
information acquisition. \citet{LuYe2013} study
private information costs and revenue-optimal entry
with efficient continuation;
\citet{LuYeFeng2021} examine shortlisting with
private initial valuation estimates; and
\citet{GolrezaeiNazerzadeh2017} show that optimal
learning with private initial information can
require weighted second-price allocation.
Related research studies endogenous information
structures, reserves, and participation
\citep{BP07,Shi12,CSZ09}, seller disclosure
\citep{GanuzaPenalva2019,GuoHaoShi2025},
and flexible information acquisition
\citep{KimKoh2022}.
Our distinguishing feature is that both valuation
priors and learning-cost scales are private, while
the seller jointly designs learning and continuation.
The resulting cross-prior constraints affect not
only information rents and experiment assignments,
but also the optimal allocation rule after learning.

The efficiency benchmark is related to VCG with
endogenous acquisition \citep{BV02} and dynamic
pivot mechanisms \citep{BV2010}.
Other related work examines learning under
interdependent values \citep{BSV09}, auction
formats and information acquisition
\citep{Persico00}, flexible attention
\citep{Mensch22}, and costly actions affecting
valuations \citep{GMSZ21}. Our analysis also
draws on cyclical monotonicity, multidimensional
screening, and dynamic envelope methods
\citep{Rochet1987,RochetChone1998,PST14}.

Section~\ref{sec:environment} introduces the
environment and mechanisms.
Section~\ref{sec:full-seller} derives the revenue
representation and private-prior incentive
characterization.
Section~\ref{subsec:reduced-mechanisms} establishes
attainment, finite-LP approximation, and joint
optimality conditions.
Section~\ref{sec:example} presents the two-prior
illustration, and Section~\ref{sec:conclusion}
concludes. Proofs appear in the Appendix, with
additional calculations in the Supplementary Appendix.

\section{Environment and mechanisms}\label{sec:environment}

\subsection{Types and values}

There are $n\geq2$ risk-neutral bidders, indexed by $i\in N=\{1,\ldots,n\}$, and one indivisible object. Bidder $i$ has an underlying payoff
\(V_i\in\{a,b\},\; 0\leq a<b.\)
Her two private stage-$1$ type components are $s_i\in[0,1]$ and $r_i\in[0,1]$; write $\theta_i=(s_i,r_i)$. For a complete type profile $\theta=(\theta_i)_{i\in N}$, write $s=(s_i)_{i\in N}$ and $r=(r_i)_{i\in N}$, so that $\theta$ may equivalently be represented as $(s,r)$.
Conditional on $s_i$,
\(
\Pr(V_i=b\mid s_i)=s_i\) and
\(\Pr(V_i=a\mid s_i)=1-s_i.
\)
For any prior $s$, define its mean payoff by
\(\mu_s:=a+s(b-a).\)
The cost scale $r_i$ affects only learning costs and, conditional on $s_i$, not $V_i$.

Preferences are quasilinear and bidders are risk neutral. If $X_i\in\{0,1\}$ denotes the realized allocation, $P_i$ the bidder's payment to the designer, and $\pi_i$ her assigned technology, realized utility is
\(X_iV_i-P_i-r_iC(\pi_i).\)
Conditional on a posterior mean $t\in[a,b]$ and an allocation probability $p_i$, the corresponding expected consumption payoff is $p_i t$. 

Payments are signed and may be negative (subsidies). The outside option is zero, so interim participation requires nonnegative truthful expected utility.

The pairs $(s_i,r_i)$ are i.i.d. across bidders with product distribution $F_s\times F_r$. We use $F_r(r)$ for the cost CDF and $F_r(dr)$ for its induced probability measure; thus $F_r(dr)=f_r(r)\,dr$. Write
\(
F:=F_s\times F_r\),
\(F_i:=F\),
\(F^n:=\bigotimes_{i=1}^n F_i\), and
\(F_{-i}:=\bigotimes_{j\neq i}F_j
\)
for the common individual, profile, and opponents' type distributions.
The prior distribution $F_s$ is an arbitrary Borel probability measure on $[0,1]$; it may be continuous, discrete, or mixed. 

Let $\mathcal S:=\operatorname{supp}(F_s)$ and take the effective type space to be the compact set
\(\Theta:=\mathcal S\times[0,1].\)
Thus incentive constraints are imposed on types in the support rather than on zero-probability prior values outside it. 
The cost distribution $F_r$ is absolutely continuous with continuous density $f_r$ that is strictly positive on $[0,1]$. 
Independence of \(s_i\) and \(r_i\) within bidder is imposed to preserve a transparent decomposition of rents, while independence across bidders is used in the conditional-averaging, screening, and revenue-approximation arguments below. The paper maintains these independence assumptions throughout. 

For each $s\in[0,1]$, let
\(\nu_s:=(1-s)\delta_a+s\delta_b.\)
Let $\mathbb P$ denote the joint distribution of types and underlying values, given by 
\begin{equation}\label{eq:joint-types-values}
\mathbb P(d\theta,dv)
=
F^n(d\theta)\prod_{i=1}^n\nu_{s_i}(dv_i).
\end{equation}
Equivalently, conditional on the full type profile $\theta$, the underlying values are independent across bidders, with $V_i\sim\nu_{s_i}$. In particular, conditional on $s_i$, the underlying value $V_i$ is independent of $r_i$, opponents' types, and opponents' underlying values. Types are observed privately at stage $1$.

\subsection{Primitive signal technologies}

Fix a finite signal space $Y=\{1,\ldots,J\}$, \(J\geq 2\). An experiment or signal technology is a stochastic kernel
\(\pi:Y\times\{a,b\}\to[0,1], \; \sum_{y\in Y}\pi(y\mid v)=1 \;\text{for }v\in\{a,b\}.\)
Identify each kernel $\pi$ with the vector $(\pi(y\mid a),\pi(y\mid b))_{y\in Y}$ in the finite-dimensional product of probability simplexes and equip the technology space with the induced Euclidean topology. Let $\Pi$ be a nonempty compact set of such technologies. Importantly, $\Pi$ does not depend on the bidder's reported prior. The assigned technology is maintained as the bidder's exclusive source of additional information about $V_i$ between the two stages: bidders cannot supplement, replace, or privately augment the assigned experiment with an outside signal. 

\citet{GP23} study when mechanisms give agents incentives to acquire information about others' preferences; the maintained restriction to assigned own-value experiments deliberately rules out that additional learning margin in order to isolate screening in private priors and learning costs. Accordingly, throughout the paper, enforceability of a signal assignment includes both conduct of the assigned technology and exclusion of unmodeled supplemental learning.

For true prior $s$, experiment $\pi$, and signal $y$, let
\(m_s^{\pi}(y) = (1-s)\pi(y\mid a)+s\pi(y\mid b)\)
be the signal probability. Whenever $m_s^{\pi}(y)>0$, define the posterior expected value
\begin{equation}\label{eq:posterior-value}
t(s,\pi,y)
=
\frac{a(1-s)\pi(y\mid a)+bs\pi(y\mid b)}{m_s^{\pi}(y)}.
\end{equation}
At zero-probability signals set $t(s,\pi,y)=a$. This convention extends $t$ to a Borel-measurable function on the whole domain and has no effect on any on-path probability or expected payoff. If $Y_s^{\pi}$ denotes the signal generated by $\pi$, then
\(T(s,\pi):=t(s,\pi,Y_s^{\pi})\)
is the posterior expected value and satisfies Bayes plausibility,
\(E[T(s,\pi)]=\mu_s.\)
It is useful to define the posterior numerator
\(n_s^{\pi}(y) = a(1-s)\pi(y\mid a)+bs\pi(y\mid b).\)
For a profile $\pi=(\pi_1,\ldots,\pi_n)\in\Pi^n$, signal realizations are conditionally independent across bidders given $(V,s,\pi)$. Conditional on $(s,\pi)$, let $T_i=T(s_i,\pi_i)$ denote bidder $i$'s posterior value.

The designer's experiment-assignment randomization is independent of underlying values conditional on the stage-$1$ reports. Given valuations and assigned experiments, bidders'
signals are independent across bidders and of the
auxiliary randomization. These independence conventions are part of the maintained generative interpretation of the mechanism and experiments.

Bidder $i$ incurs learning cost $r_iC(\pi_i)$, where
$C:\Pi\to\mathbb R_+$ is continuous and nonnegative. Since $\Pi$ is compact, define 
\(\bar C:=\max_{\pi\in\Pi}C(\pi)<\infty.\) Thus the baseline cleanly separates the two private type components: $s$ affects beliefs about value, whereas $r$ is the private multiplicative scale of a common base learning-cost function. This restriction isolates the private-prior channel.

We assume that $\Pi$ contains an uninformative technology $\pi^0$ satisfying \(\pi^0(y\mid a)=\pi^0(y\mid b)\)  for every \(y\in Y\), with $C(\pi^0)=0$. Hence $T(s,\pi^0)=\mu_s$ almost surely.
The assigned experiment generates $Y_i\sim\pi_i(\cdot\mid V_i)$; the designer sends no other pre-signal information.

\subsection{Gross surplus and the timing}

For a prior profile $s$ and experiment profile $\pi$, define
expected efficient gross surplus by
\begin{equation}\label{eq:gross-surplus}
G(s,\pi)
:=
E\left[\max_{i\in N}T_i\mid s,\pi\right].
\end{equation}
This quantity depends on the prior and experiment profiles,
but not on learning-cost scales. The
corresponding welfare benchmark deducts actual information-acquisition
costs from this gross surplus.

The mechanism has two stages. In stage $1$, each bidder privately observes $\theta_i=(s_i,r_i)$ and reports a type. The mechanism assigns a signal technology to each bidder. The assigned experiments are subsequently conducted and bidder $i$ privately observes her signal $Y_i$.

In stage $2$, bidders report their signal labels, and the continuation mechanism allocates the object and specifies payments. Payments are signed positively when paid by a bidder to the seller; transfers do not enter welfare, and the seller has zero value for retaining the object.
Unless stated otherwise, accepting the stage-$1$ mechanism commits a bidder to the assigned technology and the stage-$2$ continuation mechanism. This commitment assumption is substantive: continuation participation is not required after each signal realization.
Requiring voluntary participation after signals are observed would add history-by-history continuation-participation constraints; that alternative is not imposed here.
We restrict attention to simultaneous experimentation.
We abstract from adaptive experiment ordering and stopping based on earlier signals.

\subsection{Admissible coordinated mechanisms}\label{subsec:general-direct}

We take as primitive the class of direct coordinated mechanisms
\(M=(\Lambda,p,P)\), with no exogenous bound on payments. After stage-$1$
reports \(\widehat\theta=(\widehat s,\widehat r)\), the Borel probability
kernel \(\Lambda(d\pi\mid\widehat\theta)\) assigns the experiment profile
\(\pi\in\Pi^n\). Bidder $i$ observes her own assigned experiment $\pi_i$
and then her private signal $Y_i$. At stage $2$ she submits a signal
label \(\widehat y_i\in Y\). The Borel rules
\(p_i(\widehat\theta,\pi,\widehat y)\) and
\(P_i(\widehat\theta,\pi,\widehat y)\) determine her allocation
probability and signed payment. These rules may depend jointly on
\emph{all} bidders' type reports, assigned experiments, and signal
reports. The seller commits to the rules and may privately randomize
over assignments and allocations. Feasibility requires
\begin{equation}\label{eq:allocation-feasibility-general}
p_i^M(\widehat\theta,\pi,\widehat y)\ge0,
\qquad
\sum_i p_i^M(\widehat\theta,\pi,\widehat y)\le1.
\end{equation}

The direct signal-label report permits continuation rules to use
reported posterior values. For example, an allocation rule can
compute $t(\widehat s_i,\pi_i,\widehat y_i)$ from reports. This does
not assert a dominant-strategy VCG implementation after a false
prior report: all relevant dynamic deviations are explicitly subject
to incentive compatibility.

\paragraph{Dynamic incentives and participation.}
The private continuation history of bidder $i$ consists of her true
type $\theta_i$, own stage-$1$ report $\widehat\theta_i$, assigned
experiment $\pi_i$, and realized signal $Y_i$. Let $\Sigma_i$ be
the set of Borel strategies
\(\sigma_i:(\theta_i,\widehat\theta_i,\pi_i,Y_i)
\mapsto\widehat y_i\in Y\).
A unilateral dynamic deviation consists of a stage-$1$ report
$\widehat\theta_i$ and a strategy $\sigma_i\in\Sigma_i$, with
opponents truthful. Write $E^M_{\theta_i;\widehat\theta_i,\sigma_i}$
for its expectation, using the independent value and signal laws of
Section~\ref{sec:environment} and the assignment kernel $\Lambda$.
The truthful continuation strategy $\sigma_i^{\mathrm{tr}}$ reports
$Y_i$. The deviation payoff is
\[
U_i^M(\theta_i;\widehat\theta_i,\sigma_i)
= E^M_{\theta_i;\widehat\theta_i,\sigma_i}
\left[p_i^M V_i-P_i^M-r_iC(\pi_i)\right],
\]
where allocations and payments are evaluated at the complete induced
report and assignment profiles.

We impose deviation-wise and truthful-path absolute integrability:
\begin{equation}\label{eq:deviation-integrability}
E^M_{\theta_i;\widehat\theta_i,\sigma_i}[|P_i^M|]<\infty
\quad\text{for every }i,\theta_i,\widehat\theta_i,\sigma_i,
\end{equation}
\begin{equation}\label{eq:truthful-integrability}
E_{\mathrm{tr}}^M[|P_i^M|]<\infty
\quad\text{for every }i.
\end{equation}
Dynamic Bayesian incentive compatibility (BIC) and stage-$1$ interim
individual rationality (IR) require, pointwise,
\begin{equation}\label{eq:dynamic-BIC-general}
U_i^M(\theta_i;\theta_i,\sigma_i^{\mathrm{tr}})
\ge U_i^M(\theta_i;\widehat\theta_i,\sigma_i)
\quad\text{for all }\widehat\theta_i,\sigma_i,
\end{equation}
\begin{equation}\label{eq:interim-IR-general}
U_i^M(\theta_i;\theta_i,\sigma_i^{\mathrm{tr}})\ge0
\quad\text{for every }\theta_i.
\end{equation}
Let $\mathcal M^0$ comprise these feasible, integrable, dynamically
BIC, interim-IR mechanisms. Experiments are enforceable,
stage-$1$ acceptance binds the bidder to continuation play, and the
seller has full commitment. The revenue objective is
\[
\operatorname{Rev}(M):=E_{\mathrm{tr}}^M\left[\sum_iP_i^M\right],
\qquad R^*:=\sup_{M\in\mathcal M^0}\operatorname{Rev}(M).
\]
We use a supremum because existence of an optimal mechanism is established below for finite experiment menus, but not for general compact menus.
The seller's problem is defined over $\mathcal M^0$; we do not invoke a revelation principle for arbitrary indirect, history-dependent mechanisms. The payment-bounding reduction below applies within this maintained class.

\section{Revenue and private-prior incentives}\label{sec:full-seller}

We now express seller revenue using the learning-cost envelope
and identify how privately known priors constrain the rents left
to bidders. The revenue identity allows arbitrary continuation
allocations and experiments and requires no finite prior support.
The subsequent difference-constraint formulation specializes
only the prior support, leaving the cost scale continuous.

\subsection{The cost-type envelope}

For any direct two-stage mechanism \(M\) satisfying the feasibility and integrability requirements above, define bidder \(i\)'s truthful interim expected base learning cost pointwise from the stage-\(1\) assignment kernel by
\[
q_i^M(s,r)
:=
\int_{\Theta_{-i}}
\int_{\Pi^n}
C(\pi_i)\,
\Lambda^M(d\pi\mid(s,r),\theta_{-i})\,
F_{-i}(d\theta_{-i}),
\qquad
(s,r)\in\Theta.
\]
Thus \(q_i^M(s,r)\) is the expected base learning cost induced when bidder \(i\) truthfully reports \((s,r)\) and all opponents report truthfully. Because it is defined directly from the assignment kernel, \(q_i^M(s,r)\) is specified at every \((s,r)\in\Theta\), including \(F_i\)-null types.
The expected learning cost of type \((s,r)\) is therefore
\(r q_i^M(s,r)\). As shorthand for the truthful diagonal of the deviation payoff defined above, let
\(
U_i^M(s,r)
:=
U_i^M\bigl((s,r);(s,r),\sigma_i^{\mathrm{tr}}\bigr).
\)

We say that \(M\) satisfies \emph{continuation obedience} if, for every bidder \(i\), type \(\theta_i\), and continuation strategy \(\sigma_i\in\Sigma_i\),
\[
U_i^M\bigl(\theta_i;\theta_i,\sigma_i^{\mathrm{tr}}\bigr)
\geq
U_i^M\bigl(\theta_i;\theta_i,\sigma_i\bigr).
\]
Thus, conditional on truthful stage-\(1\) reporting, truthful continuation play is optimal against every unilateral continuation deviation. In the reduction and approximation arguments below, continuation obedience is retained explicitly for every true type and admissible continuation strategy.

We say that \(M\) satisfies \emph{within-prior dynamic IC} if, for every bidder \(i\), prior \(s\in\mathcal S\), true cost \(r\in[0,1]\), alternative cost report \(r'\in[0,1]\), and continuation strategy \(\sigma_i\in\Sigma_i\),
\begin{equation}\label{eq:within-prior-dynamic-IC}
U_i^M(s,r)
\geq
U_i^M\bigl((s,r);(s,r'),\sigma_i\bigr).
\end{equation}
Thus the bidder may change the reported cost while keeping the prior report fixed and may also choose an arbitrary continuation deviation. Taking \(r'=r\) shows that within-prior dynamic IC includes truthful continuation obedience. Below, ``within-prior \(r\)-IC'' is shorthand for the condition (\ref{eq:within-prior-dynamic-IC}).

The next lemma establishes the one-dimensional cost-type envelope for
the class $\mathcal M^0$. It requires only
within-prior incentive compatibility, not cross-prior IC, and
provides the basis for the seller's revenue representation.

\begin{lem}
\label{lem:general-r-envelope}
Let \(M\) be any direct two-stage mechanism satisfying the primitive feasibility and integrability requirements and within-prior dynamic IC \eqref{eq:within-prior-dynamic-IC}. Then, for every bidder \(i\) and every \(s\in\mathcal S\), the map
\(
q_i^M(s,r)\)
is weakly decreasing in \(r\in[0,1]
\). Moreover, \(U_i^M(s,\cdot)\) is absolutely continuous and satisfies
\begin{equation}\label{eq:general-r-envelope}
U_i^M(s,r)
=
K_i^M(s)
+
\int_r^1 q_i^M(s,z)\,dz,
\qquad
K_i^M(s):=U_i^M(s,1).
\end{equation}
If \(M\) also satisfies stage-$1$ interim IR, then
\(
K_i^M(s)\geq0
\;
\text{for every }s\in\mathcal S.
\)
\end{lem}

Every \(M\in\mathcal M^0\) satisfies within-prior dynamic IC,
and the lemma also applies to the normalized within-prior designs
used below. It isolates the part of the multidimensional problem that
remains one-dimensional: a deviation that changes only the reported
cost scale leaves the bidder's true $r$ entering utility solely
through $-rC(\pi_i)$. Hence the same envelope logic survives
conditional on $s$, despite the freedom to design continuation
allocation and payments.

\subsection{Revenue representation}

Under truthful play, conditional independence of bidders' values and
signals and the fact that assignment randomization precedes learning imply
\begin{equation}\label{eq:own-value-CI}
V_i\;\perp\!\!\!\perp\; (r_i,\theta_{-i},\pi_{-i}^M,Y_{-i})
\;\bigm|\;(s_i,\pi_i^M,Y_i).
\end{equation}
Hence, with $T_i=t(s_i,\pi_i^M,Y_i)$, iterated expectations give
$E_{\mathrm{tr}}^M[p_i^M V_i]=E_{\mathrm{tr}}^M[p_i^M T_i]$.
Revenue is therefore expected allocated posterior value less actual
learning costs and bidders' truthful utilities. By
Lemma~\ref{lem:general-r-envelope}, averaging the cost-type envelope
first over $r$ and then over $s$ gives
\[
E[U_i^M(s_i,r_i)]
=E[K_i^M(s_i)]
+E_{\mathrm{tr}}^M\!\left[
\frac{F_r(r_i)}{f_r(r_i)}C(\pi_i^M)\right].
\]
This follows by interchanging the envelope integral and expectation;
independence of prior and cost types identifies the second term.
Thus the information rent from cost screening adds
$F_r(r_i)/f_r(r_i)$ to the actual cost scale $r_i$.
Set $\psi(r)=r+F_r(r)/f_r(r)$. The resulting identity is:
\begin{prop}\label{prop:general-revenue}
For every \(M\in\mathcal M^0\),
\begin{equation}\label{eq:full-revenue-decomposition}
\operatorname{Rev}(M)
=
E_{\mathrm{tr}}^M
\left[
\sum_i
p_i^M(\theta,\pi^M,Y)T_i
-
\sum_i
\psi(r_i)C(\pi_i^M)
\right]
-
\sum_i
\int_{\mathcal S}
K_i^M(s)\,F_s(ds).
\end{equation}
\end{prop}

The representation applies to all feasible continuation
allocations and integrable signed payments in $\mathcal M^0$.
It reveals why experimentation and allocation cannot generally be
optimized separately: changing a continuation rule affects both
allocated posterior value and cross-prior rent requirements. With
efficient VCG continuation fixed, the first term is instead efficient
gross surplus $G(s,\pi)$. This paper studies the additional allocation
and payment instruments.

\subsection{Finite-prior screening and difference constraints}\label{subsec:finite-prior-optimal-design}

In this section and in the finite-prior results that follow, suppose
\(\mathcal S=\{s^{(1)},\ldots,s^{(N_S)}\}\)
with $N_S\ge2$, and write $\lambda_s:=F_s(\{s\})>0$. Cost types remain continuously distributed.
A \emph{normalized within-prior design} $\widetilde M$
has the same allocation, assignment, measurability and deviation-wise
integrability requirements as a member of $\mathcal M^0$, satisfies
within-prior dynamic IC and has $U_i^{\widetilde M}(s,1)=0$ for
every $i,s$; cross-prior IC is not required. Let $\mathcal N^0$
denote this \emph{unbounded-payment} class.

By Lemma~\ref{lem:general-r-envelope}, its truthful utility is
\(
U_i^{\widetilde M}(s,r)=\int_r^1q_i^{\widetilde M}(s,t)\,dt\ge0.
\)
At this stage no payment bound is imposed, and the class
$\mathcal N^0$ is defined independently of the bounded-payment
normalization needed for the approximation theorem.

For a normalized design define its virtual surplus and signed cross-prior deviation gains by
\begin{align*}
\mathcal V(\widetilde M)
&:=E_{\mathrm{tr}}^{\widetilde M}\left[\sum_i p_i^{\widetilde M}T_i-\sum_i\psi(r_i)C(\pi_i^{\widetilde M})\right],\\
D_{ss'}^i(\widetilde M)
&:=\sup_{r,r'\in[0,1],\,\sigma_i\in\Sigma_i}
\left\{U_i^{\widetilde M}((s,r);(s',r'),\sigma_i)-U_i^{\widetilde M}(s,r)\right\},\quad s\ne s'.
\end{align*}
The gains are real and bounded above by $b+\bar C$ (Lemma~\ref{lem:finite-prior-gain-bound}). Restoring a prior-specific boundary rent $K_{i,s}\ge0$ means \emph{reducing} every payment after a report with prior $s$ by $K_{i,s}$. The result is fully dynamically BIC exactly when
\begin{equation}\label{eq:cross-prior-economic}
K_{i,s}-K_{i,s'}\ge D_{ss'}^i(\widetilde M),\qquad i\in N,\ s\ne s'.
\end{equation}
Restoration leaves within-prior IC and continuation obedience unchanged, and the restored truthful utility is nonnegative. Revenue is $\mathcal V(\widetilde M)-\sum_{i,s}\lambda_sK_{i,s}$.

Let
\(
\mathbf K
:=
(K_{i,s})_{i\in N,\;s\in\mathcal S}
\)
denote the vector of prior-specific boundary rents; throughout, $\mathbf K\ge0$ means $K_{i,s}\ge0$ for all $i,s$. 

\begin{prop}\label{prop:finite-prior-supremum}
With finite prior support, the primitive seller's revenue supremum is
\begin{equation}\label{eq:finite-prior-economic-primal}
R^*=\sup_{\substack{\widetilde M\in\mathcal N^0,\ \mathbf K\ge0\\
K_{i,s}-K_{i,s'}\ge D_{ss'}^i(\widetilde M)\ \forall i,\ s\ne s'}}
\left\{\mathcal V(\widetilde M)-\sum_{i,s}\lambda_sK_{i,s}\right\}.
\end{equation}
\end{prop}

Every feasible pair on the right yields a primitive dynamically BIC and interim-IR mechanism. The finite-prior problem therefore separates the within-prior screening problem from the remaining cross-prior incentive constraints: the former is summarized by the envelope, while the latter reduce to finitely many directed difference constraints on \(\mathbf K\).

For a fixed $\widetilde M$, the difference constraints \eqref{eq:cross-prior-economic} admit nonnegative boundary rents if and only if every directed cycle of distinct prior classes has nonpositive total gain:
\begin{equation}\label{eq:finite-cycle-condition}
\sum_{\ell=1}^kD_{s^{(j_\ell)}s^{(j_{\ell+1})}}^i(\widetilde M)\le0,
\qquad j_{k+1}=j_1,\quad k\ge2.
\end{equation}
When feasible, the revenue-minimizing rents for a fixed normalized design solve a finite linear program (Lemma~\ref{lem:finite-difference-constraints}).

\section{Attainment and constructive approximation}\label{subsec:reduced-mechanisms}
\subsection{Attainment for finite experiment menus}

Although our mechanism domain permits integrable but unbounded
payments, a uniform bound can be imposed without reducing its
revenue supremum. The proof of the next lemma is given in
Appendix~\ref{app:bounded-reduced-payments}.

Put
\(
B_0:=5b+2\bar C.
\)
Let $\mathcal M^0(B_0)$ be the subclass of $\mathcal M^0$
with $|P_i|\le B_0$ pointwise for all bidders, reports, experiments,
and signal-report profiles. Assignments and continuation rules
may still depend on every stage-$1$ report and experiment assignment.
For the finite-prior approximation argument, also put
$B_1:=b+2B_0$ and define
$\mathcal N^0(B_1):=\{\widetilde M\in\mathcal N^0:
|P_i^{\widetilde M}|\le B_1\text{ for all }i\}$.
This is a bounded \emph{technical subclass}, not the domain of the
finite-prior revenue characterization.

\begin{lem}\label{lem:bounded-reduced-payments}
For any $M\in\mathcal M^0$ there is
$M^{\mathrm{bd}}\in\mathcal M^0(B_0)$ with the same truthful
experiment and allocation distributions and
$\operatorname{Rev}(M^{\mathrm{bd}})\ge\operatorname{Rev}(M)$.
Consequently, for arbitrary prior distributions and compact
experiment menus,
\begin{equation}\label{eq:reduced-supremum}
R^*=\sup_{M\in\mathcal M^0(B_0)}\operatorname{Rev}(M).
\end{equation}
\end{lem}

The construction preserves truthful payoffs up to a common
participation-normalizing increase in payments and never raises
the gain from a unilateral dynamic deviation.

\medskip

We next turn from equality of revenue suprema to attainment. Although the prior distribution $F_s$ may be any Borel distribution on $[0,1]$, with effective prior space $\mathcal S=\operatorname{supp}(F_s)$, and the learning-cost scale remains continuously distributed, finite experiment and signal spaces yield a compact polytope of bounded primitive outcome coordinates. A limiting kernel can be completed at exceptional type reports through a compact interim-section maximization and measurable continuation relabeling.

\begin{thm}\label{thm:direct-attainment}
Suppose $\Pi$ is finite and maintain the assumptions of
Sections~\ref{sec:environment} and~\ref{sec:full-seller},
including finite $Y$ and bounded values. Then there exists
$M^*\in\mathcal M^0(B_0)$ such that
\begin{equation}\label{eq:direct-attainment}
R^*
=\operatorname{Rev}(M^*)
=\max_{M\in\mathcal M^0(B_0)}\operatorname{Rev}(M).
\end{equation}
\end{thm}

The proof in Appendix~\ref{app:direct-attainment} combines compactness of bounded outcome kernels with measurable completion at exceptional reports. The result concerns $\mathcal M^0$ and finite $\Pi$ and $Y$, not arbitrary indirect mechanisms or infinite experiment menus.

\subsection{Constructive approximation with continuous cost types}\label{subsec:finite-LP-main}

We now suppose that both the prior support $\mathcal S$
and the experiment menu $\Pi$ are finite, with
$|\mathcal S|\ge2$. Under these restrictions, the
seller's optimal revenue admits a sequence of
finite-dimensional linear-programming approximations
that retain exact dynamic BIC and interim IR at
\emph{every} cost type.
Let $0=t_0<\cdots<t_m=1$ be a cost-report partition,
and define its maximal interval length by
\(
\delta_m:=\max_{1\le k\le m}(t_k-t_{k-1}).
\)
Consider a sequence of such partitions with $\delta_m\to0$. Let
\(B_0=5b+2\bar C\) be the continuation-payment bound
for the finite LP. Let $\mathrm{LP}_m(B_0)$ denote a finite linear program defined by \eqref{lp:feas}--\eqref{lp:obj} in
Appendix~\ref{app:finite-experiment-LP}.

\begin{thm}\label{thm:finite-experiment-LP}
Suppose $\mathcal S$ and $\Pi$ are finite with
$|\mathcal S|\ge2$.
There are finite linear programs $\mathrm{LP}_m(B_0)$
whose feasible points induce finite-category primitive
mechanisms satisfying exact dynamic BIC and interim IR
for every $r\in[0,1]$.
Their optimal revenues $R_m^{\mathrm{LP}}(B_0)$ converge
to $R^*$ at a rate proportional to the maximal interval
length $\delta_m$, with the model primitives and finite
prior support held fixed.
\end{thm}

Although the LP uses finitely many cost categories, it
enforces exact dynamic BIC and interim IR for every cost
type. Within each category, truthful utility and the payoff
difference against any fixed deviation are affine in the
actual cost. Since the LP includes all relevant continuation-report
strategies, the endpoint constraints
\eqref{lp:ic}--\eqref{lp:ir} are necessary and sufficient.
Moreover, the transfer-normalization lemma in
Appendix~\ref{app:finite-experiment-LP} establishes that
$\mathrm{LP}_m(B')$ and $\mathrm{LP}_m(B_0)$ have the same
optimal value for every finite $B'\ge B_0$. Thus, the
payment bound $B_0$ entails no loss relative to any larger
finite bound, even for a fixed partition.

The convergence proof in Appendix~\ref{app:finite-experiment-LP}
starts from an arbitrary admissible mechanism, applies an
incentive-preserving cell-envelope adjustment, and eliminates
residual cross-prior incentive violations by mixing with a
strict benchmark. The argument neither requires continuity of
the virtual cost scale $\psi$ nor relies on
Theorem~\ref{thm:direct-attainment}, which separately
establishes attainment of $R^*$ for finite experiment menus.

\paragraph{Joint finite-LP optimality conditions.}
In efficient continuation, allocation follows posterior values. For revenue, the shadow values of incentive and participation constraints modify both continuation allocation and the relative attractiveness of experiments. The LP dual expresses these effects as rent-adjusted scores and provides joint optimality conditions for assignment, allocation, and payments.

Let $h_i=(\widehat s_i,k_i)\in H_m:=\mathcal S\times\{1,\ldots,m\}$ denote bidder $i$'s
reported prior and cost category and let $h=(h_1,\ldots,h_n)\in H_m^n$ be a complete
stage-$1$ report profile.
For each experiment profile $\pi\in\Pi^n$, let
$\gamma_{h,\pi}=\Pr(\pi\mid h)$ denote its assignment probability
conditional on $h$. The weighted continuation
allocation and payment variables are
\(
\xi_{i,h,\pi,\widehat y}
=\gamma_{h,\pi}\,p_i(\widehat y\mid h,\pi)\) and 
\(\zeta_{i,h,\pi,\widehat y}
=\gamma_{h,\pi}\,P_i(\widehat y\mid h,\pi)
\), respectively,
where $\widehat y\in Y^n$ is the continuation-report
profile, and $p_i$ and $P_i$ denote bidder $i$'s
allocation probability and payment, respectively.

Let $\mathcal R_m(\zeta)$ be the linear revenue objective in \eqref{lp:obj}; although it depends directly only on weighted payments, feasibility jointly constrains $(\gamma,\xi,\zeta)$.
Write $\mathrm{ICslack}$ and $\mathrm{IRslack}$
for the left-minus-right sides of the endpoint
incentive and participation constraints
\eqref{lp:ic} and \eqref{lp:ir}, respectively.
Let $\beta_{i,s,k,j;h_i',\sigma_i}\ge0$ and
$\eta_{i,s,k,j}\ge0$ be their corresponding
multipliers, where $j\in\{k-1,k\}$.

For fixed multipliers, the Lagrangian can be
written as
\begin{align*}
\mathcal L(\gamma,\xi,\zeta;\beta,\eta)
&=\mathcal R_m(\zeta)
  +\sum\beta\,\mathrm{ICslack}
  +\sum\eta\,\mathrm{IRslack}\\
&=\sum_{h,\pi}a_{h,\pi}\gamma_{h,\pi}
 +\sum_{i,h,\pi,\widehat y}
 c_{i,h,\pi,\widehat y}\,\xi_{i,h,\pi,\widehat y}\\
&\quad+\sum_{i,h,\pi,\widehat y}
 d_{i,h,\pi,\widehat y}\,\zeta_{i,h,\pi,\widehat y}.
\end{align*}
Here $a_{h,\pi}$ collects the learning-cost
terms weighted by the incentive multipliers,
while $c_{i,h,\pi,\widehat y}$ are the
\emph{rent-adjusted allocation coefficients}
and $d_{i,h,\pi,\widehat y}$ are the coefficients
of the weighted continuation-payment variables.
The rent adjustment incorporates the shadow values
of dynamic incentive and participation constraints,
not necessarily a fixed deduction from posterior values.
These coefficients are given
explicitly in Appendix~\ref{app:finite-experiment-LP}.

Define the \emph{rent-adjusted experiment score}
\begin{equation}\label{eq:joint-lp-experiment-score}
\mathcal J_h(\pi;\beta,\eta)
:=a_{h,\pi}+\sum_{\widehat y\in Y^n}
\left[
\max\{0,\max_i c_{i,h,\pi,\widehat y}\}
+B_0\sum_i|d_{i,h,\pi,\widehat y}|
\right].
\end{equation}
The three components capture learning costs, continuation allocation, and bounded payments, each adjusted for incentive and participation constraints. Because the multipliers are endogenous, this is an optimality condition rather than a closed-form rule in primitives.

\begin{thm}\label{thm:joint-finite-LP-rules}
Under the assumptions of
Theorem~\ref{thm:finite-experiment-LP},
the finite program $\mathrm{LP}_m(B_0)$ satisfies
\begin{equation}\label{eq:joint-lp-value-dual}
R_m^{\mathrm{LP}}(B_0)
=\min_{\beta,\eta\ge0}\sum_{h\in H_m^n}
\max_{\pi\in\Pi^n}\mathcal J_h(\pi;\beta,\eta),
\end{equation}
with attainment on both sides.
A feasible LP solution \((\gamma,\xi,\zeta)\) is optimal if and only if there exist nonnegative multipliers \((\beta,\eta)\) such that:
\begin{enumerate}
\item[(i)]
$\gamma_{h,\pi}>0$ only if
$\pi\in\arg\max_{\pi'}
\mathcal J_h(\pi';\beta,\eta)$;

\item[(ii)]
whenever $\gamma_{h,\pi}>0$, its continuation
allocation $p_i(\widehat y\mid h,\pi)$ at each $\widehat y$ maximizes
$\sum_i c_{i,h,\pi,\widehat y}\,p_i$
over $p_i\ge0$, $\sum_i p_i\le1$;

\item[(iii)]
on the same branches,
$P_i(\widehat y\mid h,\pi)=B_0$
if $d_{i,h,\pi,\widehat y}>0$, and
$P_i(\widehat y\mid h,\pi)=-B_0$
if $d_{i,h,\pi,\widehat y}<0$;
when $d_{i,h,\pi,\widehat y}=0$,
any payment in $[-B_0,B_0]$ is permitted
by this pointwise condition;

\item[(iv)]
all endpoint IC and IR inequalities hold,
and their slacks vanish whenever their
respective multipliers are strictly positive.
\end{enumerate}
\end{thm}

Under a positively assigned experiment, continuation allocates only to bidders with maximal nonnegative rent-adjusted coefficients; if all coefficients are negative, the object is withheld. These coefficients incorporate the shadow values of dynamic incentive and participation constraints rather than only posterior valuations. 

Experiment mixing and payments must still satisfy IC and IR. In part (iii), the extreme-payment rule reflects the bound $B_0$; when $d=0$, IC and IR determine payments, if at all. The LP jointly determines $(\gamma,\xi,\zeta)$, and $P_i=\zeta_i/\gamma_{h,\pi}$ on positive-probability branches.

\section{Economic implications: a two-prior illustration}\label{sec:example}\label{par:two-prior-example}

The two-prior example compares efficient with revenue-optimal continuation and fixed with cost-contingent experimentation. The Supplementary Appendix establishes the revenue bounds and verifies incentive compatibility and participation for $\hat M$.

Consider two bidders with $V_i\in\{0,1\}$,
equally likely private priors
$s_i\in\{L,H\}$ where $L=1/5$ and $H=2/5$, and independent
$r_i\sim U[0,1]$. Each bidder can receive
either a free uninformative experiment $\pi^0$
or an informative binary experiment $\pi^1$,
with
$\Pr(+\mid V_i=1)=6/7$,
$\Pr(+\mid V_i=0)=1/7$,
and base cost $C(\pi^1)=1/245$.

First consider continuation. The \emph{posterior-bid VCG benchmark},
with unrestricted posterior bids in stage~$2$, earns at most
$7/20=0.35$ even under coordinated experiments. Its reporting
space is richer than the primitive signal-label space, so this
bound does not apply to every efficient signal-label mechanism.
By contrast, informing both bidders, subsidizing each by $1/245$,
and running a second-price auction with reserve $3/5$ yields a
dynamically BIC, interim-IR revenue of $173/490\approx0.35306$.
The reserve is Myerson-optimal for the unconditional posterior
distribution \citep{Myerson81}; the prior report and signal label
jointly encode a posterior bid.

With both informative experiments fixed but
continuation allocation and payments unrestricted,
optimal revenue is
\(
R_{\mathrm{fixed}}^*=\frac{19}{49}
\), approximately $0.38776$.
The optimal continuation allocates according
to the largest strictly positive
\emph{rent-adjusted value} $\phi_{s,y}$, using
\[
\begin{array}{c|cc}
 & Y=- & Y=+ \\ \hline
L & -3/25 & 3/10 \\
H & 1/10 & 4/5
\end{array}
\]
and withholds the object if all rent-adjusted values
are nonpositive.

These scores account for cross-prior rents: allocations after
low-prior reports are discounted to deter high-prior imitation.
Thus the seller withholds the object when two low-prior bidders
receive negative signals, despite positive posterior values,
whereas two high-prior negative signals still induce allocation.
The Supplementary Appendix proves that \emph{any} efficient
continuation with fixed experiments earns at most
$73/196<19/49$. This strict inefficiency result holds within
$\mathcal M^0$ and is distinct from the posterior-bid VCG
benchmark. The scores illustrate the rent-adjusted allocation
principle in Theorem~\ref{thm:joint-finite-LP-rules}.

Allowing correlated experiment assignments contingent on
reported priors, but not reported costs, leaves optimal
revenue unchanged, even with general continuation rules:
\(
R_{\mathrm{CI}}^*
=R_{\mathrm{fixed}}^*
=\frac{19}{49}.
\)

Conditioning experimentation on reported learning
costs, however, yields a strict improvement.
To see how, consider an explicit mechanism
$\hat M$ with two cost-report categories,
separated at $r=1/2$. Its experiment-assignment
rule is as follows. Both bidders receive
$\pi^1$ whenever at least one reports prior $L$.
When both report prior $H$, both receive $\pi^1$
if both report the lower-cost category.
Otherwise, only one bidder is informed:
the bidder reporting the lower-cost category,
with uniform randomization when the categories
coincide.

When both bidders are informed, continuation
allocation follows the rent-adjusted ranking
above. When only one bidder is informed,
she receives the object after a positive signal,
while the uninformed bidder receives it otherwise.
Appropriate continuation payments make this
mechanism dynamically BIC and interim IR.
Its revenue is
\begin{equation}
\label{eq:two-prior-endogenous-revenue}
\operatorname{Rev}(\hat M)
=\frac{3043}{7840}
=R_{\mathrm{CI}}^*+\frac{3}{7840}
>R_{\mathrm{CI}}^*
=R_{\mathrm{fixed}}^*.
\end{equation}

Learning matters differently across priors. For a low-prior
bidder, the signal separates positive from negative rent-adjusted
values. With two high-prior bidders, both signal outcomes have
positive scores; informing a second bidder improves the choice
of winner but adds learning costs and rents. The rule therefore
informs both high-prior bidders only when both report low costs.
This is the tradeoff reflected in the rent-adjusted experiment scores of Theorem~\ref{thm:joint-finite-LP-rules}.

\section{Conclusion}\label{sec:conclusion}

Private priors and private learning costs interact through both
experimentation and continuation design. Our revenue identity
separates allocated posterior value, virtual learning costs, and
prior-dependent boundary rents. Cross-prior constraints may
prevent a normalized design from being implementable or require
additional rents; learning and continuation therefore cannot
generally be optimized independently across prior classes.

We establish revenue attainment for finite experiment menus and
arbitrary prior distributions. With finite prior support, incentive
constraints reduce to directed difference inequalities; finite
linear programs then approximate optimal revenue while preserving
exact dynamic IC and interim IR at every continuous cost type.
Their dual characterizes jointly optimal assignments, allocations,
and payments. The two-prior example illustrates both inefficient
optimal continuation and gains from cost-contingent learning.

Extensions include constructive approximations with continuum
prior distributions and more primitive characterizations of optimal
rent-adjusted assignments. Sequential experiments, nonenforceable
learning, and weaker commitment features would introduce further incentive
and information-design considerations.

\appendix
\section{Proofs}\label{app:proofs}

\subsection{The cost-type envelope}

\paragraph{General cost-type envelope.}

\begin{proof}[Proof of Lemma \ref{lem:general-r-envelope}]
Fix bidder $i$ and $s\in\mathcal S$. Let $r<r'$. Consider
deviations that keep the reported prior equal to $s$, change the
reported cost scale, and then truthfully report $Y_i$ at stage $2$.
Conditional on the reported cost, the induced assignment, signal,
allocation, and payment distributions are independent of the bidder's
true cost; $r$ enters utility only through $-rC(\pi_i)$.

Applying within-prior dynamic IC in both directions, between reports $(s,r)$ and $(s,r')$, gives the pairwise bounds
\begin{equation}\label{eq:general-r-envelope-pairwise}
(r'-r)q_i^M(s,r')
\leq
U_i^M(s,r)-U_i^M(s,r')
\leq
(r'-r)q_i^M(s,r).
\end{equation}
It follows immediately that $q_i^M(s,\cdot)$ is weakly decreasing. Since $0\leq q_i^M(s,r)\leq\bar C$, \eqref{eq:general-r-envelope-pairwise} also makes $U_i^M(s,\cdot)$ Lipschitz and therefore absolutely continuous. At almost every $r$, the derivative satisfies
$\partial U_i^M(s,r)/\partial r=-q_i^M(s,r)$.
Integrating from $r$ to $1$ and setting $K_i^M(s)=U_i^M(s,1)$ gives \eqref{eq:general-r-envelope}. Finally, stage-$1$ interim IR at $(s,1)$ gives $K_i^M(s)\geq0$.
\end{proof}

\subsection{Finite-prior constraints and revenue characterization}
\label{app:finite-prior-gains}
\label{app:finite-prior-supremum-proof}

\paragraph{Finite difference constraints.}

The following elementary graph result characterizes feasibility of the finite-prior boundary-rent system used in the general joint-design problem.

\begin{lem}\label{lem:finite-difference-constraints}
Let \(\mathcal S\) be finite and let \(d_{ss'}\in\mathbb R\) be given for every ordered pair \(s\neq s'\). The system
\(
K_s-K_{s'}\geq d_{ss'}
\;(s\neq s'),
\;
K_s\geq0
\quad(s\in\mathcal S)
\)
is feasible if and only if every directed cycle
\(s_1\to\cdots\to s_k\to s_1\)
has nonpositive total weight,
\(
d_{s_1s_2}+\cdots+d_{s_ks_1}\leq0.
\)
Whenever the system is feasible and \(\lambda_s>0\) for every \(s\), the problem
\[
\min_{\mathbf K\geq0}
\left\{\sum_s\lambda_sK_s:
K_s-K_{s'}\geq d_{ss'}\ \forall s\neq s'\right\}
\]
attains its minimum.
\end{lem}

\begin{proof}
Necessity follows by summing the difference constraints around a directed cycle, which cancels the \(K\)-terms. For sufficiency, form the reversed constraint graph: for each economic edge \(s\to s'\) of weight \(d_{ss'}\), place a computational edge \(s'\to s\) with the same weight. 
Adjoin an artificial source \(o\) with a zero-weight edge \(o\to s\) to every node. If every directed cycle has nonpositive total weight, the maximum weight of a directed path from \(o\) to any \(s\) is finite, because cycles of nonpositive weight can be deleted and it is enough to consider simple paths. Let \(K_s\) equal this maximum path weight. The edge \(o\to s\) gives \(K_s\geq0\), and the edge \(s'\to s\) of weight \(d_{ss'}\) gives
\(
K_s\geq K_{s'}+d_{ss'},
\)
which is the desired difference constraint.
When the system is feasible, its feasible set is a nonempty closed polyhedron in the nonnegative orthant. If every \(\lambda_s>0\), each sublevel set of \(\sum_s\lambda_sK_s\) is bounded. Hence the weighted minimum is attained.
\end{proof}

\paragraph{Finite cross-prior deviation gains.}

The following bound ensures that cross-prior deviation gains remain finite even when the original mechanism permits unrestricted signed transfers.
\begin{lem}\label{lem:finite-prior-gain-bound}
For every $\widetilde M\in\mathcal N^0$, bidder $i$,
and distinct prior classes $s,s'$,
\[-\infty<D_{ss'}^i(\widetilde M)\le b+\bar C.\]
\end{lem}

\begin{proof}
Fix $i$, priors $s\ne s'$, and a deviation by true type $(s,r)$ to report $(s',r')$ with continuation strategy $\sigma_i$. Compare it with the same stage-$1$ report by hypothetical type $(s',1)$. The latter can ignore her own experiment signal and, conditional on her observed pre-signal history (including $\pi_i$), privately simulate an independent signal with distribution $m_s^{\pi_i}$, then apply the original strategy $\sigma_i$ to the simulated history with the constants $(s,r)$ inserted in place of her actual type. 

Since experiment assignments precede signals and depend only on
reports, and opponents' types and values are independent of bidder
$i$'s true type, the joint distributions of the simulated continuation
report, payments, and allocation probabilities coincide with those
under the original deviation. Their expected allocated-value terms
may differ, but each belongs to $[0,b]$. Changing the cost scale
from $r$ to $1$ changes utility by at most $\bar C$.
Hence the original deviation payoff is no greater than the
simulated payoff of type $(s',1)$ plus $b+\bar C$.

If the simulation uses private randomization, it is a mixture of deterministic Borel continuation strategies. The maintained deviation-wise absolute integrability and equality of the simulated and original distributions permit integration over the random seed; within-prior IC bounds the payoff of every deterministic strategy in this mixture by $U_i^{\widetilde M}(s',1)=0$. Thus the original deviation payoff is at most $b+\bar C$. The cost envelope and boundary normalization give \(U_i^{\widetilde M}(s,r)=\int_r^1q_i^{\widetilde M}(s,z)\,dz\ge0.\) Taking the supremum proves the upper bound. Every individual deviation has a finite payoff by deviation-wise integrability, and at least one deviation is available, so the supremum is a real number.
\end{proof}

\paragraph{Proof of the finite-prior revenue characterization.}

The characterization is an algebraic equivalence in the
unbounded-payment classes $\mathcal M^0$ and $\mathcal N^0$.
No compactness or uniform payment bound is needed.
\begin{proof}[Proof of Proposition~\ref{prop:finite-prior-supremum}]
Take any $M\in\mathcal M^0$ and put
$K_{i,s}:=U_i^M(s,1)\ge0$. For every report with prior $s$,
\emph{increase} bidder $i$'s continuation payment by $K_{i,s}$.
The resulting $\widetilde M$ has boundary utility zero, preserves
within-prior dynamic IC and truthful continuation obedience, and
belongs to $\mathcal N^0$. Since $\mathcal S$ is finite, every
$K_{i,s}$ is finite and the report-contingent payment shifts preserve
deviation-wise and truthful-path absolute integrability. Restoring
the rent vector reproduces $M$, so full dynamic BIC implies
\eqref{eq:cross-prior-economic}.

By Lemma~\ref{lem:general-r-envelope} and the revenue decomposition
\eqref{eq:full-revenue-decomposition},
\[
\operatorname{Rev}(M)
=
E_{\mathrm{tr}}^{M}
\left[
\sum_i p_i^{M}T_i
-
\sum_i\psi(r_i)C(\pi_i^{M})
\right]
-
\sum_{i,s}\lambda_sK_{i,s}.
\]
The normalization changes only continuation payments, leaving
the truthful distributions of allocations and assigned experiments
unchanged. Hence the expectation equals
$\mathcal V(\widetilde M)$, and therefore
\[
\operatorname{Rev}(M)
=
\mathcal V(\widetilde M)
-
\sum_{i,s}\lambda_sK_{i,s}.
\]
Thus the supremum on the right of \eqref{eq:finite-prior-economic-primal} is at least $R^*$.

Conversely, take any right-hand feasible pair. Reduce every
continuation payment after a report with prior $s$ by $K_{i,s}$.
Within-prior IC is unchanged. The inequalities
\eqref{eq:cross-prior-economic} deter all cross-prior stage-$1$ and
continuation deviations; the cost envelope and $K_{i,s}\ge0$ give
interim IR. Since the prior set is finite, these finite shifts
preserve both integrability requirements. The resulting member of
$\mathcal M^0$ yields the displayed objective. Neither direction
requires an optimizer.
\end{proof}

\subsection{Bounded payments within the reduced class}
\label{app:bounded-reduced-payments}

The following argument constructs bounded continuation payments
within $\mathcal M^0$ while preserving dynamic incentives.

\begin{proof}[Proof of Lemma~\ref{lem:bounded-reduced-payments}]
Fix $M\in\mathcal M^0$ and bidder $i$. All opponents report
truthfully. Conditional on $i$'s fixed stage-$1$ report
$\widehat\theta_i$, the joint law of the assigned own experiment
$\pi_i$ and the hidden variables
$H_{-i}:=(\theta_{-i},\pi_{-i},Y_{-i})$ is independent of bidder
$i$'s unreported true type. This follows from independence of types,
the report-contingent assignment kernel, and the independent
experimental signals. Therefore, a jointly Borel disintegration gives
a conditional law of $H_{-i}$ given
$(\widehat\theta_i,\pi_i)$, uniformly in the parameter
$\widehat\theta_i$.

For each possible own continuation report $z\in Y$, let
\[
x_z(\widehat\theta_i,\pi_i)
:=E[p_i^M(\widehat\theta,\pi,z,Y_{-i})
       \mid\widehat\theta_i,\pi_i],\qquad
h_z(\widehat\theta_i,\pi_i)
:=E[P_i^M(\widehat\theta,\pi,z,Y_{-i})
       \mid\widehat\theta_i,\pi_i].
\]
The expectation integrates out truthful opponents and their signals.
For any fixed $\widehat\theta_i$, deviation-wise absolute
integrability for the strategy always reporting $z$ makes each
$h_z$ finite outside a null set of assigned $\pi_i$; finiteness
of $Y$ gives a common null set. Choose jointly Borel versions and
complete these functions on the report-dependent null sets. Those
sets are null after the same report for \emph{every} true type, since
the distribution of $\pi_i$ depends on the report, not the true type.

Call a label $y\in Y$ \emph{active} under
$(\widehat s_i,\pi_i)$ if
$m_{\widehat s_i}^{\pi_i}(y)>0$.
For every active label $y$, truthful continuation
obedience at type $\widehat\theta_i$ implies,
for every label $z$,
\[
t(\widehat s_i,\pi_i,y)x_y-h_y
\ge t(\widehat s_i,\pi_i,y)x_z-h_z
\quad\text{for almost every assigned }\pi_i.
\]
Otherwise the bidder could deviate only after receiving $y$ on
the Borel set of experiments where the inequality fails. Taking
both inequalities for two active labels $y,z$ gives
$|h_y-h_z|\le b$, because $0\le x_y,x_z\le1$ and
$0\le t(\widehat s_i,\pi_i,\cdot)\le b$.
These inequalities can be completed on report-contingent null sets
without altering truthful outcomes or any stage-$1$ deviation value.

For each $(\widehat s_i,\pi_i)$, choose measurably the smallest
\emph{active} label $y^0$ with
$m_{\widehat s_i}^{\pi_i}(y^0)>0$ and put
$v_i(\widehat\theta_i,\pi_i):=h_{y^0}$. On the active labels
replace bidder $i$'s continuation payment by
\[
P_i^{\mathrm{res}}(\widehat\theta,\pi,\widehat y)
:=h_{\widehat y_i}(\widehat\theta_i,\pi_i)
       -v_i(\widehat\theta_i,\pi_i).
\]
This payment is bounded in absolute value by $b$. For an inactive
label $z$ under $(\widehat s_i,\pi_i)$, set bidder $i$'s allocation
to zero and payment to $b$. In the residual rule the reference
label $y^0$ gives conditional payoff
$t(s_i,\pi_i,y_i)x_{y^0}\ge0$ for \emph{any} true prior, while
an inactive report gives $-b$. Consequently, an inactive report is
dominated by an active one, including after a false prior report.
Dropping inactive reports cannot raise the best continuation
deviation payoff. This alteration of bidder $i$'s allocation only
removes a nonnegative allocation coordinate and preserves feasibility;
it does not change outcomes with bidder $i$ truthful.

Let
\(
\eta_i(\widehat\theta_i)
:=E[v_i(\widehat\theta_i,\pi_i)
       \mid\widehat\theta_i].
\)
It is finite by the always-reference-label deviation and
deviation-wise integrability. For any true type, own stage-$1$
report, and continuation strategy using active labels, conditional
independence of hidden variables from the own signal implies that
the expected payoff under the residual rule, \emph{minus}
$\eta_i(\widehat\theta_i)$, equals the original payoff.
Thus one may temporarily interpret $\eta_i$ as an own-report
stage-$1$ fee. The transformation preserves truthful interim
utilities, expected revenue, and all dynamic incentive constraints,
and cannot raise the value of a cross-prior or within-prior
deviation. Moreover, truthful-path absolute integrability follows
from $|h_{Y_i}-v_i|\le b$ and the original payment integrability.

For this temporary representation let
$J_i(\theta_i;\widehat\theta_i)$ be the maximal expected payoff
over continuation strategies \emph{excluding} $\eta_i$. Since
the residual payment lies in $[-b,b]$, allocated value in $[0,b]$,
and learning cost in $[0,\bar C]$,
\[
L:=-(b+\bar C)\le J_i(\theta_i;\widehat\theta_i)
\le H:=2b.
\]
Full dynamic BIC, applied in both directions between any two
types, gives
$|\eta_i(\theta_i)-\eta_i(\widehat\theta_i)|\le H-L$.
Put $\delta_i:=\inf_{\theta_i}U_i^M(\theta_i)\ge0$ and raise
every own-report fee by $\delta_i$. This preserves dynamic BIC
and interim IR and weakly raises revenue; truthful utilities now
have infimum zero. Writing $\eta_i$ for the adjusted fee, IR gives
$\eta_i(\theta_i)\le H$ for all reports. Choosing a sequence of
true types with truthful utility tending to zero and using the
fee-difference bound gives $\eta_i(\widehat\theta_i)\ge2L-H$.
Hence
\(
-4b-2\bar C\le\eta_i(\widehat\theta_i)\le2b.
\)

Finally, add this own-report fee to every residual continuation
payment, leaving no stage-$1$ fee. The resulting payment satisfies
$|P_i|\le b+4b+2\bar C=B_0$, and total utility and incentives
are unchanged. Repeat independently for each bidder. Every step
changes only that bidder's payment and, on her off-grid reports,
possibly removes her allocation, so feasibility and other bidders'
unilateral incentives are preserved. We obtain
$M^{\mathrm{bd}}\in\mathcal M^0(B_0)$ with weakly higher revenue.
\end{proof}

\subsection{Attainment with finite experiment menus}\label{app:direct-attainment}

We prove Theorem~\ref{thm:direct-attainment} by taking strongly
convergent convex averages of bounded outcome kernels and using
weak-$*$ compactness of interim sections to complete the limit
at exceptional reports.

Let $Z$ be standard Borel with Borel probability $\nu$.
For bounded measurable $f:Z\to\mathbb R^d$, write
$[f]\in L^\infty(\nu;\mathbb R^d)$ for its $\nu$-a.e.
equivalence class. Given nonempty compact convex
$D\subset\mathbb R^d$, put
\(
\mathcal K_D
=
\{k\in L^\infty(\nu;\mathbb R^d):
k(z)\in D\ \nu\text{-a.e.}\},
\)
equipped with the Borel structure induced by
$\sigma(L^\infty,L^1)$.

\begin{lem}\label{lem:attainment-borel-representative}
Let $X,Z$ be standard Borel spaces, $\nu$ a Borel probability measure on $Z$, and $D\subset\mathbb R^d$ a nonempty compact convex set. If $x\mapsto k_x\in\mathcal K_D$ is weak-$*$ Borel, there exists a jointly Borel map $F:X\times Z\to D$ such that
$[F(x,\cdot)]=k_x$ in $L^\infty(\nu;\mathbb R^d)$ for \emph{every} $x\in X$.
\end{lem}
\begin{proof}
Choose increasing finite Borel partitions $(\mathcal P_q)$
generating, up to $\nu$-null sets,
\(
\sigma\!\left(\bigcup_{q\geq1}\mathcal P_q\right)
=\mathcal B(Z)\).
Fix $d^0\in D$. For each cell $A\in\mathcal P_q$ with $\nu(A)>0$, set
\(
 a_A(x):=\frac{1}{\nu(A)}\int_Ak_x(z)\,\nu(dz),
\)
interpreting the integral by weak-$*$ duality. Each $a_A$
is Borel and $D$-valued. For null cells put $a_A=d^0$,
and let $F_q(x,z)=a_A(x)$ for $z\in A$. For each fixed $x$
this is a conditional expectation of $k_x$ on $\mathcal P_q$.
Bounded martingale convergence gives $F_q(x,\cdot)\to k_x$
almost everywhere. Set $F$ equal to this limit on its Borel
convergence set and to $d^0$ elsewhere. Closedness of $D$
gives the asserted jointly Borel representative.
\end{proof}

\begin{proof}[Proof of Theorem~\ref{thm:direct-attainment}]
\emph{Step 1: a finite-dimensional bounded outcome kernel.}

Let $\Theta=\mathcal S\times[0,1]$, and write $\mu=F_s\otimes F_r$. For each complete first-stage report profile $\widehat\theta\in\Theta^n$, experiment profile $\pi\in\Pi^n$ and continuation signal-report profile $\widehat y\in Y^n$, set
\[
\gamma_\pi(\widehat\theta):=\Pr(\pi\mid \widehat\theta),\qquad
\xi_{i,\pi,\widehat y}(\widehat\theta):=\gamma_\pi(\widehat\theta)p_i(\widehat y\mid \widehat\theta,\pi),\qquad
\zeta_{i,\pi,\widehat y}(\widehat\theta):=\gamma_\pi(\widehat\theta)P_i(\widehat y\mid \widehat\theta,\pi).
\]
Since $\Pi$ and $Y$ are finite, these coordinates comprise a vector $\Gamma(\widehat\theta)=(\gamma,\xi,\zeta)(\widehat\theta)\in\mathbb R^d$, for a fixed finite $d$. Feasibility and the transfer bound are equivalent to
\begin{equation}\label{eq:attainment-polytope}
\sum_\pi\gamma_\pi=1,\quad\gamma_\pi\ge0,\quad
\xi_{i,\pi,\widehat y}\ge0,\quad\sum_i\xi_{i,\pi,\widehat y}\le\gamma_\pi,\quad
-B_0\gamma_\pi\le\zeta_{i,\pi,\widehat y}\le B_0\gamma_\pi.
\end{equation}
Let $D\subset\mathbb R^d$ be the compact convex polytope defined by \eqref{eq:attainment-polytope}. Every Borel map $\Gamma:\Theta^n\to D$ implements a primitive reduced mechanism: use the conditional rules $p_i=\xi_i/\gamma_\pi$ and $P_i=\zeta_i/\gamma_\pi$ if $\gamma_\pi>0$, and choose zero allocation and payment otherwise. The recovered rules are Borel and feasible, and $|P_i|\le B_0$.

For bidder $i$, actual type $(s,r)$, first-stage report $\widehat\theta_i$, and a pure continuation strategy $\sigma:\Pi\times Y\to Y$, let $\Phi_i(s,r;\widehat\theta_i,\sigma,\Gamma)$ denote her interim expected utility against truthful opponents. Its explicit expression is
\begin{align}\label{eq:attainment-deviation-functional}
\Phi_i(s,r;\widehat\theta_i,\sigma,\Gamma)
={}&\int_{\Theta_{-i}}\sum_{\pi\in\Pi^n}\sum_{y\in Y^n}
\left(\prod_{j\ne i}m_{s_j}^{\pi_j}(y_j)\right)
\Bigl[n_s^{\pi_i}(y_i)\xi_{i,\pi,(\sigma(\pi_i,y_i),y_{-i})}(\widehat\theta_i,\theta_{-i})\notag\\
&\hspace{28mm}-m_s^{\pi_i}(y_i)\zeta_{i,\pi,(\sigma(\pi_i,y_i),y_{-i})}(\widehat\theta_i,\theta_{-i})\Bigr]\,\mu^{n-1}(d\theta_{-i})\notag\\
&-r\int_{\Theta_{-i}}\sum_{\pi\in\Pi^n}C(\pi_i)\gamma_\pi(\widehat\theta_i,\theta_{-i})\,\mu^{n-1}(d\theta_{-i}).
\end{align}
Here $m_s^\pi$ and $n_s^\pi$ are the signal probabilities and
posterior numerators defined in Section~\ref{sec:environment}.
The integrand is bounded; payoff is linear in the section
$\Gamma(\widehat\theta_i,\cdot)$ and affine in $r$.
In the primitive class $\mathcal M^0$, continuation strategies
report only signal labels and observe no additional designer messages.
Thus, for fixed $(\theta_i,\widehat\theta_i)$, a pure continuation
strategy can be represented as $\sigma:\Pi\times Y\to Y$.
Since $\Pi,Y$ are finite, there are finitely many such strategies;
randomization cannot improve on their best payoff.

\emph{Step 2: a strongly convergent maximizing sequence.}
By \eqref{eq:reduced-supremum}, choose Borel reduced kernels $\Gamma^k:\Theta^n\to D$ with $\operatorname{Rev}(\Gamma^k)\to R^*$. Since $D$ is bounded, $(\Gamma^k)$ is bounded in the Hilbert space $L^2(\mu^n;\mathbb R^d)$. Passing to a subsequence gives weak convergence to some $\Gamma^0$. Because $D$ is closed and convex, $\Gamma^0(\widehat\theta)\in D$ for $\mu^n$-almost every $\widehat\theta$. Choose a Borel $D$-valued version everywhere.

By Mazur's lemma, finite convex combinations of tails,
$\bar\Gamma^k$, satisfy
\begin{equation}\label{eq:attainment-strong-convergence}
\|\bar\Gamma^k-\Gamma^0\|_{L^2(\mu^n)}\le2^{-k}.
\end{equation}
Because payoff comparisons and revenue are affine in weighted
outcomes, each $\bar\Gamma^k$ remains dynamically BIC and interim
IR, with $\operatorname{Rev}(\bar\Gamma^k)\to R^*$.
Truthful revenue is a bounded linear functional on $L^2$; hence
\begin{equation}\label{eq:attainment-limit-revenue}
\operatorname{Rev}(\Gamma^0)=\lim_k\operatorname{Rev}(\bar\Gamma^k)=R^*.
\end{equation}

\emph{Step 3: uniform interim utilities and ordinary-report incentive constraints.}
For each bidder $i$, write $U_i^k(s,r)$ for truthful interim utility under $\bar\Gamma^k$. By interim IR and bounded payments,
\[
0\le U_i^k(s,r)\le b+B_0\] for all \((s,r)\in\Theta.
\)
The affine dependence of $m_s^\pi,n_s^\pi$ on $s$ in
\eqref{eq:attainment-deviation-functional} gives, uniformly
in reports, plans, and $k$,
\begin{equation}\label{eq:attainment-Lipschitz-payoff}
|\Phi_i(s,r;\widehat\theta_i,\sigma,\bar\Gamma^k)-\Phi_i(s',r';\widehat\theta_i,\sigma,\bar\Gamma^k)|
\le L_s|s-s'|+\bar C|r-r'|,
\qquad L_s:=2(b+B_0).
\end{equation}
The respective allocation and payment changes sum to at most
$2b|s-s'|$ and $2B_0|s-s'|$. By dynamic BIC,
\(
U_i^k(s,r)=\sup_{\widehat\theta_i\in\Theta,\,\sigma\in\Sigma_i^0}\Phi_i(s,r;\widehat\theta_i,\sigma,\bar\Gamma^k).
\)
The supremum is equi-Lipschitz on compact
$\Theta=\mathcal S\times[0,1]$. Arzel\`a--Ascoli gives
continuous $U_i:\Theta\to\mathbb R_+$ and a common subsequence with
\begin{equation}\label{eq:attainment-utility-uniform}
\max_i\sup_{(s,r)\in\Theta}|U_i^k(s,r)-U_i(s,r)|\longrightarrow0.
\end{equation}

The summable $L^2$ estimate in \eqref{eq:attainment-strong-convergence}, Fubini, and Cauchy--Schwarz supply, for each bidder $i$, a Borel set $E_i\subseteq\Theta$ with $\mu(E_i)=1$, such that for every report $\widehat\theta_i\in E_i$,
\begin{equation}\label{eq:attainment-section-convergence}
\int_{\Theta_{-i}}\|\bar\Gamma^k(\widehat\theta_i,\theta_{-i})-\Gamma^0(\widehat\theta_i,\theta_{-i})\|\,\mu^{n-1}(d\theta_{-i})\longrightarrow0.
\end{equation}
Choose $E_i$ so that, for every actual type $(s,r)\in\Theta$, every $\widehat\theta_i\in E_i$, and every pure continuation plan $\sigma$, passage to the limit in the dynamic IC inequalities gives
\begin{equation}\label{eq:attainment-ordinary-IC}
\Phi_i(s,r;\widehat\theta_i,\sigma,\Gamma^0)\le U_i(s,r).
\end{equation}
For $\widehat\theta_i=(s,r)\in E_i$, truthful continuation achieves
\begin{equation}\label{eq:attainment-ordinary-truth}
\Phi_i(s,r;(s,r),\mathrm{id},\Gamma^0)=U_i(s,r).
\end{equation}
The latter equality holds for ordinary types, while \eqref{eq:attainment-ordinary-IC} controls deviations from \emph{every} true prior and cost type into every ordinary report.

\emph{Step 4: compact interim sections and payoff-maximizing completion.}
For bidder $i$, collect the feasible sections of $\Gamma^0$
as equivalence classes in the weak-$*$ compact metric space
\[
\mathcal K_i:=\{k\in L^\infty(\mu^{n-1};\mathbb R^d):k(\theta_{-i})\in D\;\mu^{n-1}\text{-a.e.}\}.
\]
Linear inequalities defining $D$ are weak-$*$ closed; Banach--Alaoglu
and separability of $L^1(\mu^{n-1})$ give compact metrizability.
For $\widehat\theta_i\in E_i$, the section map
$k_i(\widehat\theta_i):=[\Gamma^0(\widehat\theta_i,\cdot)]$
is weak-$*$ Borel by testing against a countable determining family.

Let $k^0\in\mathcal K_i$ assign the free uninformative
experiment to everyone, with zero allocation and payments.
Complete the ordinary-section menu by taking
\begin{equation}\label{eq:attainment-ordinary-section-closure}
\mathcal K_i^{\mathrm{ord}}
:=
\operatorname{cl}_{w^*}
\bigl(k_i(E_i)\cup\{k^0\}\bigr)
\subseteq\mathcal K_i,
\end{equation}
where $\operatorname{cl}_{w^*}$ denotes weak-$*$ closure.
For a section $k$ define $\Phi_i(s,r;k,\sigma)$ by the right side of \eqref{eq:attainment-deviation-functional} with $\Gamma(\widehat\theta_i,\cdot)$ replaced by $k$, and define
\begin{equation}\label{eq:attainment-envelope-sections}
V_i(s,r):=\max_{k\in\mathcal K_i^{\mathrm{ord}},\ \sigma\in\Sigma_i^0}\Phi_i(s,r;k,\sigma),
\qquad \Sigma_i^0:=\{\sigma:\Pi\times Y\to Y\}.
\end{equation}
For fixed $(s,r,\sigma)$ this payoff is weak-$*$ continuous in $k$, because the integrands are bounded measurable functions of $\theta_{-i}$; it is jointly continuous in $(s,r,k)$ by \eqref{eq:attainment-Lipschitz-payoff}. Compactness and finiteness of $\Sigma_i^0$ ensure the maximum is attained and the correspondence of maximizers admits a Borel selection
\(
(s,r)\longmapsto(k_i^*(s,r),\sigma_i^*(s,r)).
\)
The function $V_i$ is continuous. By \eqref{eq:attainment-ordinary-IC}, the payoff from every ordinary section and continuation plan is bounded above by $U_i(s,r)$ for \emph{any} actual type $(s,r)$. Weak-$*$ continuity extends the bound to the closure. The zero section also yields $0\le U_i(s,r)$, hence $V_i(s,r)\le U_i(s,r)$ everywhere. For an ordinary actual type $\theta_i\in E_i$, its own section and truthful continuation attain $U_i(\theta_i)$ by \eqref{eq:attainment-ordinary-truth}, so
\begin{equation}\label{eq:attainment-ordinary-value}
V_i(\theta_i)=U_i(\theta_i)\qquad(\theta_i\in E_i).
\end{equation}

For an exceptional stage-$1$ report $\widehat\theta_i=(s,r)\notin E_i$, select the maximizing section $k_i^*(\widehat\theta_i)$ and plan $\sigma_i^*(\widehat\theta_i)$. The relabeling is applied to the \emph{entire outcome vector}, not only bidder $i$'s allocation or payment. In the coordinates of \eqref{eq:attainment-polytope}, for a fixed plan $\sigma\in\Sigma_i^0$ define $R_{i,\sigma}:D\to D$ by
\begin{align}\label{eq:attainment-full-coordinate-relabeling}
 (R_{i,\sigma}z)^\gamma_\pi&=z^\gamma_\pi,\notag\\
 (R_{i,\sigma}z)^\xi_{j,\pi,\widehat y}
 &=z^\xi_{j,\pi,(\sigma(\pi_i,\widehat y_i),\widehat y_{-i})},
 &&j\in N,\notag\\
 (R_{i,\sigma}z)^\zeta_{j,\pi,\widehat y}
 &=z^\zeta_{j,\pi,(\sigma(\pi_i,\widehat y_i),\widehat y_{-i})},
 &&j\in N.
\end{align}
Each reported-label row is copied from an entire feasible outcome row, so the single-object constraint and every payment bound are preserved, even when $\sigma$ is not one-to-one. The map $R_{i,\sigma}$ is linear and continuous and induces a weak-$*$ continuous map on feasible sections. Define
\(k_i^\dagger(\widehat\theta_i)
 :=R_{i,\sigma_i^*(\widehat\theta_i)}k_i^*(\widehat\theta_i).\)
This section map is weak-$*$ Borel, since $\Sigma_i^0$ is finite and the maximizing selection is Borel. Truthful continuation in the relabeled section implements the maximizing plan, so
\begin{equation}\label{eq:attainment-null-truth}
\Phi_i(\widehat\theta_i;k_i^\dagger(\widehat\theta_i),\mathrm{id})=V_i(\widehat\theta_i).
\end{equation}
For any actual type $\theta_i'$ and any continuation plan $\rho\in\Sigma_i^0$, the full-coordinate identity \eqref{eq:attainment-full-coordinate-relabeling} gives
\(
\Phi_i(\theta_i';k_i^\dagger(\widehat\theta_i),\rho)
 =\Phi_i(\theta_i';k_i^*(\widehat\theta_i),
                \sigma_i^*(\widehat\theta_i)\circ\rho),
\)
where \[(\sigma_i^*\circ\rho)(\pi_i,y_i)
=\sigma_i^*(\widehat\theta_i)(\pi_i,\rho(\pi_i,y_i))\] is again in $\Sigma_i^0$. It follows that
\begin{equation}\label{eq:attainment-null-IC}
\Phi_i(\theta_i';k_i^\dagger(\widehat\theta_i),\rho)\le V_i(\theta_i')\qquad\text{for all actual }\theta_i'.
\end{equation}
In particular, relabeling creates no profitable deviations relative to the compact-menu value $V_i$.

\emph{Step 5: equality at all types and a single measurable primitive mechanism.}
We first show that $V_i$ and $U_i$ coincide \emph{everywhere}. They agree on $E_i$ by \eqref{eq:attainment-ordinary-value}. This set has full $\mu$-measure and is dense in \(\Theta=\operatorname{supp}(F_s)\times[0,1],\) since $F_r$ has full support. Both functions are continuous by \eqref{eq:attainment-utility-uniform} and \eqref{eq:attainment-envelope-sections}. Therefore
\begin{equation}\label{eq:attainment-value-equality}
V_i(s,r)=U_i(s,r)\ge0\qquad\text{for every }(s,r)\in\Theta.
\end{equation}
Combining \eqref{eq:attainment-ordinary-IC}, \eqref{eq:attainment-ordinary-truth}, \eqref{eq:attainment-null-truth}, \eqref{eq:attainment-null-IC}, and \eqref{eq:attainment-value-equality}, we have the requisite truthful payoff and the bound on each deviation at \emph{every} report, once the chosen interim sections are implemented in one primitive mechanism.

Apply Lemma~\ref{lem:attainment-borel-representative} with $X=\Theta$, $Z=\Theta_{-i}$, $\nu=\mu^{n-1}$, and the polytope $D$ from \eqref{eq:attainment-polytope}. Extend the weak-$*$ Borel map $\widehat\theta_i\mapsto k_i^\dagger(\widehat\theta_i)$ from $\Theta\setminus E_i$ to all of $\Theta$ by assigning the constant section $k^0$ on $E_i$. The lemma yields a jointly Borel, everywhere $D$-valued function $\Gamma_i^\dagger(\widehat\theta_i,\theta_{-i})$ representing the selected section for \emph{each fixed} exceptional $\widehat\theta_i$, modulo a $\mu^{n-1}$-null set that may depend on that report. Such report-dependent null sets cause no problem in interim payoff comparisons.

Set $\Gamma^*(\widehat\theta)=\Gamma^0(\widehat\theta)$ when all reported $\widehat\theta_j$ belong to their respective ordinary sets $E_j$. Otherwise select the smallest index $i$ with $\widehat\theta_i\notin E_i$ and set
\begin{equation}\label{eq:attainment-priority}
\Gamma^*(\widehat\theta_i,\widehat\theta_{-i})=\Gamma_i^\dagger(\widehat\theta_i,\widehat\theta_{-i}).
\end{equation}
The regions on which index $i$ has priority are Borel, so this defines a single Borel $D$-valued primitive kernel. The priority rule is compatible with the \emph{unilateral} incentive constraints for the following reason. Fix \emph{any} actual type $\theta_i$ (whether or not it lies in $E_i$), any first-stage deviation $\widehat\theta_i$, and any continuation strategy $\rho$. Truthful opponents have reports in $E_{-i}:=\prod_{j\ne i}E_j$ with probability one. Hence, for $\mu^{n-1}$-almost every opponent report profile,
\[
 \Gamma^*(\widehat\theta_i,\theta_{-i})=
 \begin{cases}
 \Gamma^0(\widehat\theta_i,\theta_{-i}),
       &\widehat\theta_i\in E_i,\\
 \Gamma_i^\dagger(\widehat\theta_i,\theta_{-i}),
       &\widehat\theta_i\notin E_i.
 \end{cases}
\]
Thus \eqref{eq:attainment-ordinary-IC} or \eqref{eq:attainment-null-IC} bounds this deviation by $U_i(\theta_i)$, respectively. When $\theta_i\in E_i$, truthful continuation attains $U_i(\theta_i)$ by \eqref{eq:attainment-ordinary-truth}; when $\theta_i\notin E_i$, it attains $U_i(\theta_i)=V_i(\theta_i)$ by \eqref{eq:attainment-null-truth} and \eqref{eq:attainment-value-equality}. Profiles with two or more exceptional reports need no extra incentive argument: they occur with zero probability when all but one bidder report truthfully.
These payoff comparisons prove exact dynamic BIC and interim IR
for every actual type and every report. Randomized signal-label
continuation strategies cannot outperform the finite pure plans.
The kernel $\Gamma^*$ differs from $\Gamma^0$ only on a $\mu^n$-null set of report profiles. Therefore \eqref{eq:attainment-limit-revenue} gives
\(
\operatorname{Rev}(\Gamma^*)=\operatorname{Rev}(\Gamma^0)=R^*.
\)
The primitive implementation of $\Gamma^*$ completes the proof.
\end{proof}

\subsection{Finite-LP approximation and optimality conditions}\label{app:finite-experiment-LP}
This subsection proves Theorem~\ref{thm:finite-experiment-LP} directly from arbitrary admissible mechanisms. Assume both $\mathcal S$ and $\Pi$ are finite; costs remain continuously distributed and every incentive constraint is imposed at every actual cost type.
Fix a sequence of partitions $0=t_0<\cdots<t_m=1$, with cells
\(
I_1=[0,t_1],\;\)
\(I_k=(t_{k-1},t_k]\;\) for \(k=2,\ldots,m\), and let
\(\mathcal I_m:=\{I_1,\ldots,I_m\}.
\)
Assume every cell has positive $F_r$-probability. Define the maximal interval length by
\(
\delta_m:=\max_{1\le k\le m}(t_k-t_{k-1}),
\)
and suppose $\delta_m\to0$.
Write $B_0=5b+2\bar C$, $B_1=b+2B_0$, and fix the auxiliary
payment bound used in the approximation construction,
\begin{equation}\label{eq:lp-uniform-payment-bound}
B:=2b+3B_0+2\bar C.
\end{equation}
The finite LP itself will use the smaller bound $B_0$.

\subsubsection*{Strict cross-prior incentive benchmark} \label{app:strict-prior-benchmark}

Fix $0<\epsilon_0<1/n$. Define a normalized zero-learning design $\bar M$ by assigning $\pi^0$ to every bidder, ignoring cost reports, and assigning bidder $i$ the object with probability $\epsilon_0\widehat s_i$, charging continuation payment
\(
\bar P_i(\widehat s_i)=\epsilon_0\left[a\widehat s_i+(b-a)\widehat s_i^2\right].
\)
It is feasible, independent of signal reports, and has $|\bar P_i|\le b$. Its truthful utility is identically zero. Put
\(\bar K_{i,s}=\epsilon_0(b-a)s^2/2\).
Restoring $\bar K$ gives the payment $\epsilon_0[as+(b-a)s^2/2]$ and strictly truthful prior reporting: its truthful-minus-mimicry payoff is
\(\epsilon_0(b-a)(s-s')^2/2\).
Thus
\begin{equation}\label{eq:strict-prior-slack}
\bar K_{i,s}-\bar K_{i,s'}-D_{ss'}^i(\bar M)\ge\rho,
\qquad\rho:=\frac{\epsilon_0(b-a)}2\min_{s\ne s'}(s-s')^2>0.
\end{equation}
Denote its restored revenue by $J_0$. The pair $(\bar M,\bar{\mathbf K})$ is strictly feasible for every directed cross-prior inequality. The strictly positive margin \(\rho\) will be used to repair cross-prior incentive constraints in the finite-LP approximation.

\subsubsection*{Finite linear program}
Use the partition $\mathcal I_m$ defined above, with the stated endpoint convention. Let $H_m=\mathcal S\times\{1,\ldots,m\}$, where each report
$h_i=(\widehat s_i,k_i)\in H_m$ specifies a reported prior and a
reported cost-category index. Write \(h=(h_1,\ldots,h_n)\in H_m^n\)
for the complete stage-$1$ report profile. 
For truthful opponents, define
the report-profile probability weights
\[
w_{-i}(h_{-i})=\prod_{j\ne i}\lambda_{s_j}F_r(I_{k_j}),
\qquad
w(h)=\prod_j\lambda_{s_j}F_r(I_{k_j}),
\qquad h_j=(s_j,k_j).
\]

A \emph{finite-category primitive mechanism} is specified by an
experiment-assignment rule $\gamma_{h,\pi}$, continuation allocation
rules $p_i(\widehat y\mid h,\pi)$, and continuation payment rules
$P_i(\widehat y\mid h,\pi)$, where $\pi\in\Pi^n$ and
$\widehat y\in Y^n$. Here, $\gamma_{h,\pi}$ is the probability
of assigning experiment profile $\pi$ following the complete
stage-$1$ report profile $h$.
For the linear-programming formulation, define the weighted
allocation and payment variables
\(
\xi_{i,h,\pi,\widehat y}
:=\gamma_{h,\pi}p_i(\widehat y\mid h,\pi)\) and 
\(\zeta_{i,h,\pi,\widehat y}
:=\gamma_{h,\pi}P_i(\widehat y\mid h,\pi)
\), respectively. 
With the prescribed continuation-payment bound $B_0$, impose the linear constraints
\begin{equation}\label{lp:feas}
\gamma_{h,\pi}\ge0,\quad\sum_\pi\gamma_{h,\pi}=1,\quad
\xi_{i,h,\pi,\widehat y}\ge0,\quad\sum_i\xi_{i,h,\pi,\widehat y}\le\gamma_{h,\pi},\quad
-B_0\gamma_{h,\pi}\le\zeta_{i,h,\pi,\widehat y}\le B_0\gamma_{h,\pi}.
\end{equation}

The mechanism depends on numerical cost reports only through their
reported categories and uses the primitive signal-label continuation
rules without stage-$1$ fees or additional designer messages.
For a bidder of true type $(s,r)$, truthful stage-$1$ reporting
in this class means $(s,k_m(r))$, where $k_m(r)$ is the index
of the partition cell containing $r$; truthful continuation
means reporting $Y_i$. Throughout this subsection,
$U_i^M(s,r)$ denotes the expected utility under these
truthful reports. Unilateral deviations may involve any
alternative prior and cost-category report, together with
a continuation signal-report strategy
$\sigma_i:\Pi\times Y\to Y$.

For bidder $i$, true prior $s$, an own report $h_i'\in H_m$,
and a pure continuation strategy $\sigma_i:\Pi\times Y\to Y$,
define the continuation report profile $\widehat y^{\sigma_i}\in Y^n$ by
\[
\widehat y_j^{\sigma_i}
=
\begin{cases}
\sigma_i(\pi_i,y_i), & j=i,\\
y_j, & j\ne i.
\end{cases}
\]
Define the following linear functionals, where
$A_i(s;h_i',\sigma_i)$ is expected allocated value less continuation
payments  with $m_s^{\pi}(y)$ and $n_s^{\pi}(y)$ from Section~\ref{sec:environment}, and $Q_i(h_i')$ is expected base information cost:
\begin{align}\label{lp:Adef}
A_i(s;h_i',\sigma_i):={}&
\sum_{h_{-i}}w_{-i}(h_{-i})\sum_{\pi\in\Pi^n}\sum_{y\in Y^n}
\Bigl(\prod_{j\ne i}m_{s_j}^{\pi_j}(y_j)\Bigr)\notag\\[-2mm]
&\qquad\cdot\Bigl[
 n_s^{\pi_i}(y_i)\xi_{i,(h_i',h_{-i}),\pi,\widehat y^{\sigma_i}}
-m_s^{\pi_i}(y_i)\zeta_{i,(h_i',h_{-i}),\pi,\widehat y^{\sigma_i}}
\Bigr],\\
Q_i(h_i') :={}&\sum_{h_{-i}}w_{-i}(h_{-i})
\sum_{\pi\in\Pi^n}\gamma_{(h_i',h_{-i}),\pi}C(\pi_i).\label{lp:Qdef}
\end{align}

Thus, for a bidder of true type $(s,r)$, the expected utility from
the deviation $(h_i',\sigma_i)$ is
\[
A_i(s;h_i',\sigma_i)-rQ_i(h_i').
\]
Let $\mathrm{id}(\pi_i,y_i)=y_i$. For every $i,s,k$, every $h_i'$
and every pure $\sigma_i$, impose both endpoint inequalities
\begin{equation}\label{lp:ic}
A_i(s;(s,k),\mathrm{id})-t_jQ_i(s,k)
\ge A_i(s;h_i',\sigma_i)-t_jQ_i(h_i'),
\qquad j=k-1,k,
\end{equation}
where $Q_i(s,k):=Q_i((s,k))$. Also impose interim IR at both endpoints:
\begin{equation}\label{lp:ir}
A_i(s;(s,k),\mathrm{id})-t_jQ_i(s,k)\ge0,\qquad j=k-1,k.
\end{equation}
The objective is the following linear expression for expected truthful revenue:
\begin{equation}\label{lp:obj}
\max_{\gamma,\xi,\zeta}\;
\sum_{h\in H_m^n}w(h)
\sum_{\pi\in\Pi^n}\sum_{y\in Y^n}
\left(\prod_jm_{s_j}^{\pi_j}(y_j)\right)
\sum_i\zeta_{i,h,\pi,y}.
\end{equation}
Call this program $\mathrm{LP}_m(B_0)$ and denote its optimal value by $R_m^{\mathrm{LP}}(B_0)$.
For any other finite $B'\ge0$, let $\mathrm{LP}_m(B')$ denote
the analogous program with $B'$ replacing $B_0$ in
\eqref{lp:feas}.
For each true cost category, both truthful expected utility and its
difference from the expected utility of any fixed deviation are affine
in the bidder's actual cost $r$. Consequently, the endpoint inequalities
\eqref{lp:ic}--\eqref{lp:ir} are necessary and sufficient for dynamic
BIC and interim IR at every $r\in[0,1]$. 
Pure continuation strategies
suffice because expected utility is linear in continuation
randomization. The weighted feasibility constraints \eqref{lp:feas}
also permit reconstruction of primitive allocation and payment rules
when $\gamma_{h,\pi}>0$; when $\gamma_{h,\pi}=0$, they force the
corresponding weighted allocation and payment variables to vanish, so
an arbitrary feasible bounded continuation rule may be assigned on
that branch. Thus $\mathrm{LP}_m(B_0)$ maximizes revenue exactly over
the dynamically BIC and interim-IR members of the specified
finite-category mechanism class.
To embed any such mechanism into $\mathcal M^0$, map each
numerical stage-$1$ report $(\widehat s_i,\widehat r_i)$ to
$(\widehat s_i,k_m(\widehat r_i))$ and use the corresponding
rules $(\gamma,p,P)$, with signal-label continuation.
Every numerical-report deviation thereby induces a
category-report deviation covered by \eqref{lp:ic}, and truthful
payoffs and seller revenue are unchanged. Consequently,
$R_m^{\mathrm{LP}}(B_0)\leq R^*$.

\subsubsection*{Finite-category transfer normalization}
The normalization of Lemma~\ref{lem:bounded-reduced-payments}
preserves finite-category dependence: if the original assignment and
continuation rules depend on numerical cost reports only through
their categories, then the conditional averages, reference signal
labels, and resulting report-dependent fees constructed in its proof
depend on the same categories only.

\begin{lem}\label{lem:finite-category-transfer-normalization}
Fix $m$. Every dynamically BIC and interim-IR primitive
finite-category mechanism with bounded signed continuation
payments has a finite-category counterpart with the same experiment
assignment and truthful continuation allocations, weakly higher
revenue, and payments bounded by $B_0=5b+2\bar C$.
Consequently, for every finite $B'\ge B_0$,
\begin{equation}\label{eq:finite-category-bound-equality}
R_m^{\mathrm{LP}}(B')=R_m^{\mathrm{LP}}(B_0).
\end{equation}
\end{lem}
\begin{proof}
Apply the construction in the proof of
Lemma~\ref{lem:bounded-reduced-payments} to the category mechanism
embedded into $\mathcal M^0$. Each conditional distribution used
there is a function only of the reported prior, reported cost
category, and assigned own experiment. Consequently the output
remains a finite-category mechanism. It preserves the assignment
and truthful allocations and weakly raises revenue. Hence every
feasible point of $\mathrm{LP}_m(B')$ is dominated by a feasible
point of $\mathrm{LP}_m(B_0)$; reverse inclusion when $B'\ge B_0$
gives equality.
\end{proof}

\subsubsection*{Averaging auxiliary pre-signal randomization}

The finite-category construction below temporarily makes auxiliary
random draws visible to bidders. They can be eliminated by the
following elementary averaging fact, which is stated here only for
the auxiliary information generated by that construction.

\begin{lem}\label{lem:auxiliary-averaging}
Suppose that, conditional on the complete stage-$1$
report profile $h$ and assigned experiment profile
$\pi$, a signal-label continuation mechanism uses
auxiliary randomization $A$, with Borel conditional
law $\nu_{h,\pi}$. Assume that this law is independent
of unreported true types and that, conditional on
the true prior profile and $(h,\pi)$, $A$ is
independent of underlying valuations and experimental
signals.
Replace the continuation rules by
\[
\bar p_i(h,\pi,\widehat y)
=\int p_i(h,\pi,\widehat y,a)\,\nu_{h,\pi}(da),\quad
\bar P_i(h,\pi,\widehat y)
=\int P_i(h,\pi,\widehat y,a)\,\nu_{h,\pi}(da).
\]
If the original coordinates are bounded and feasible, then these
averaged rules remain bounded and feasible, preserve truthful
interim utility and revenue, and weakly reduce the maximal
continuation-deviation payoff after every fixed stage-$1$ report.
Thus they preserve any satisfied dynamic IC and interim IR.
\end{lem}
\begin{proof}
Fix an actual type, own stage-$1$ report, and Borel continuation
strategy in the averaged mechanism. In the auxiliary mechanism the
bidder can use the same strategy, ignoring $A$. Conditional on the
complete report and assignment profiles and actual priors, $A$
is independent of experimental outcomes, so integrating first over
$A$ gives exactly the payoff in the averaged mechanism. Taking the
supremum over continuation strategies proves the deviation
inequality. Truthful signal reporting also ignores $A$, giving
equality of truthful payments and utilities. Convexity of the
allocation simplex, boundedness and conditional integration give
the remaining assertions.
\end{proof}

\subsubsection*{Incentive-preserving cost discretization}
\begin{lem}\label{lp:approx}
Let $\widetilde M\in\mathcal N^0(B_1)$, with $\Pi$ finite.
There is a finite-category normalized mechanism $\widetilde M_m$
satisfying exact within-prior dynamic IC and continuation obedience
for every $r\in[0,1]$, with continuation payments bounded by
$B_1+2\bar C\delta_m$, such that
\begin{align}
\big|\operatorname{Rev}(\widetilde M_m)-\operatorname{Rev}(\widetilde M)\big|&\le2n\bar C \delta_m,\label{lp:revenueerr}\\
D_{ss'}^i(\widetilde M_m)&\le D_{ss'}^i(\widetilde M)+3\bar C \delta_m\quad(s\ne s').\label{lp:Derr}
\end{align}
The truthful utility for every $(s,r)$ differs by at most $\bar C \delta_m$.
\end{lem}

The proof uses fictitious within-cell reports and a
category-dependent common payment adjustment. The latter
restores exact cost-type IC, while cross-prior deviation
gains can increase by at most $3\bar C\delta_m$, uniformly
across bidders and distinct prior pairs. The revenue bound
follows from preservation of the simulated fine-report
distribution, without requiring continuity of the
virtual-cost function.

\begin{proof}

\emph{Simulation.}
For each bidder $i$ reporting $(s_i,k_i)$ at
stage $1$, the designer draws an auxiliary cost
$R_i'\sim F_r(\cdot\mid I_{k_i})$, independently
across bidders conditional on their reported
categories. The designer then runs $\widetilde M$
using the fictitious numerical report profile
$((s_i,R_i'))_{i\in N}$, with these auxiliary cost draws
serving as the cost-report inputs to the original mechanism.

Initially each bidder privately observes her own simulated cost
$R_i'$. The auxiliary mechanism (before the payment adjustment
and averaging out this simulated report) has truthful interim
learning-cost assignment
\[
q_{i,s,k}=\int_{I_k}q_{i,s}(t)\,F_r(dt\mid I_k),\qquad
q_{i,s}(t)=\mathbb E[C(\pi_i)\mid \text{truthful own report }(s,t)].
\]
Under truthful category reports, the simulated cost profile has law
$F_r^n$ and is independent of the prior profile. Following any unilateral
category report, the simulated costs of truthful opponents still have
their original distribution. Thus, conditional on a simulated own cost,
the bidder faces the original deviation environment against truthful
opponents. Actual cost $r$ enters only through the sunk learning cost
$-rC(\pi_i)$ and does not affect continuation preferences.

\emph{Envelope adjustment.} Fix $i,s$ and suppress indices. The original within-prior IC and zero boundary yield a nonincreasing $q:[0,1]\to[0,\bar C]$ and $U(r)=\int_r^1q(t)dt$, after choosing a monotone version of $q$. Define \[q^{(m)}(r)=q_k:=\mathbb E[q(R')\mid I_k] \;\text{for}\; r\in I_k, \;\text{and}\; U^{(m)}(r)=\int_r^1q^{(m)}(t)dt.\] As $q^{(m)}$ is nonincreasing, $U^{(m)}$ is continuous and convex. On $I_k$, $U^{(m)}(r)=A_k^{(m)}-rq_k$ for a constant $A_k^{(m)}$. Put \(A(t)=U(t)+tq(t) \;\text{and}\; A_k=\mathbb E[A(R')\mid I_k].\) Increase bidder $i$'s \emph{continuation payment} (uniformly across continuation reports after cost-category report $(s,k)$) by the category-specific constant
\begin{equation}\label{lp:payment-adjustment}
\delta_{i,s,k}=A_k-A_k^{(m)}.
\end{equation}
The utility from a same-prior report of cost category $k$, followed
by truthful continuation, becomes $A_k^{(m)}-rq_k$. The affine pieces
of the convex function $U^{(m)}$ are supporting lines, so
\[U^{(m)}(r)=\max_k\{A_k^{(m)}-rq_k\},\] establishing exact within-prior
cost-report IC. This assertion concerns continuation obedience for a truthful-prior report; it does not by itself establish cross-prior IC, which is controlled below by bounding the deviator's best continuation payoff. Continuation truthfulness remains optimal after each
simulated same-prior report: conditional on the experiment and signal,
actual cost $r$ is sunk, and the payment adjustment
\eqref{lp:payment-adjustment} is common to all continuation actions.
Thus continuation deviations cannot improve the payoff, and
$U^{(m)}(1)=0$.

\emph{Uniform approximation bounds.} On any $I_k$ the number $q_k$ lies between the essential infimum and supremum of $q$ on that cell. Thus, using monotonicity and $\mathrm{Var}(q)\le\bar C$,
\begin{equation}\label{lp:envelopeerr}
\sup_{r\in[0,1]}|U^{(m)}(r)-U(r)|
\le\int_0^1|q^{(m)}(t)-q(t)|dt\le \delta_m\,\mathrm{Var}(q)\le\bar C \delta_m.
\end{equation}
For $t,r\in I_k$, original IC gives
\(
0\le U(r)-[U(t)+(t-r)q(t)]\le\bar C|t-r|\le\bar C \delta_m.
\)
Averaging over $t$ in the cost category and comparing the resulting $A_k-rq_k$ with $U^{(m)}(r)$ using \eqref{lp:envelopeerr} establishes
\begin{equation}\label{lp:adjustment-bound}
|\delta_{i,s,k}|\le 2\bar C \delta_m.
\end{equation}

\emph{Cross-prior gains.} Under a false-prior cost-category report $(s',k)$, and conditional on the auxiliary cost draw $R_i'=t\in I_k$, every continuation strategy corresponds to a strategy admissible in $\widetilde M$ following exact stage-$1$ report $(s',t)$. By the definition of the cross-prior gain, its expected utility before the adjustment is at most \(U_{i,s}(r)+D_{ss'}^i(\widetilde M),\) uniformly in $t$. 
Averaging over $t$ and including the payment shift changes this upper bound by at most $2\bar C \delta_m$; subtracting the new truthful utility, which by \eqref{lp:envelopeerr} is at least $U_{i,s}(r)-\bar C \delta_m$, gives \eqref{lp:Derr}. The same argument includes arbitrary continuation deviations after false-prior reports.

\emph{Truthful revenue.} Under truthful category reporting, independent simulated fine-cost reports have precisely the original product law $F_r^n$, independently of priors. Hence, before the corrections \eqref{lp:payment-adjustment}, the induced joint law of assignments, values, signals, truthful continuation allocations and \emph{payments} is the same as in $\widetilde M$. 
The correction changes each bidder's expected continuation payment by at most $2\bar C \delta_m$ by \eqref{lp:adjustment-bound}; thus
\(
\big|\operatorname{Rev}(\widetilde M_m)-\operatorname{Rev}(\widetilde M)\big|\le 2n\bar C \delta_m.
\)
The equality of unadjusted expected payments follows from the simulated-report distribution and does not involve $\psi$.

\emph{Averaging out simulation information.} Let $h$ be the complete
reported-category profile, $\pi$ the assigned experiment profile,
and let $\mathsf A$ collect the auxiliary simulated cost draws
and any associated pre-signal randomization affecting continuation.

Conditional on $(h,\pi)$, its Borel law
$\nu_{h,\pi}$ is independent of the deviator's unreported true type;
conditional on $(h,\pi)$ and the true prior profile, $\mathsf A$
is independent of underlying values and experimental signals.
Conditioning on the assigned experiment can change the distribution
of auxiliary draws, so we use the conditional kernel given the
\emph{complete} category-report and experiment profiles $(h,\pi)$.
The original mechanism is Borel with bounded payments, so
Lemma~\ref{lem:auxiliary-averaging} applies.

The required independence is conditional on the actual prior profile
and $(h,\pi)$, not merely on $(h,\pi)$ after integrating out priors.
For each fixed true type and first-stage report, a continuation
strategy in the averaged mechanism can be replicated by ignoring
the simulated cost message; hence the lemma controls deviations,
not only truthful-path distributions.

Average the allocation and payment coordinates conditional on
$(h,\pi)$, including the payment shift \eqref{lp:payment-adjustment}.
The final mechanism reveals no simulated costs, preserves truthful
utility and revenue, and weakly reduces continuation-deviation
opportunities after every stage-$1$ report.

Thus exact within-prior IC, continuation obedience, \eqref{lp:revenueerr}, and \eqref{lp:Derr} survive. Since $H_m$, $\Pi$, and $Y$ are finite, the averaged continuation rules have finitely many coordinates. Their payments satisfy the bound in the lemma statement.
\end{proof}

\subsubsection*{Cross-prior repair and revenue convergence}
\begin{lem}\label{lp:mix}
Let $\widetilde M\in\mathcal N^0(B_1)$ and $\mathbf K$ satisfy all cross-prior constraints \eqref{eq:cross-prior-economic} and for all $i,s$, $0\le K_{i,s}\le b+B_0$. Let $\widetilde M_m$ be supplied by Lemma~\ref{lp:approx} and $(\bar M,\bar{\mathbf K})$ the strictly feasible benchmark of \eqref{eq:strict-prior-slack}. Put $d_m=3\bar C \delta_m$ and $\eta_m=d_m/(\rho+d_m)$. There exists a finite-cost-category, no-disclosure, zero-stage-$1$-fee mechanism satisfying full dynamic BIC and interim IR for every $r\in[0,1]$, having payments bounded by $B=2b+3B_0+2\bar C$ and revenue at least
\begin{equation}\label{lp:repair-rev}
(1-\eta_m)\left[\mathcal V(\widetilde M)-\sum_{i,s}\lambda_sK_{i,s}-2n\bar C \delta_m\right]
+\eta_m J_0.
\end{equation}
\end{lem}
\begin{proof}
After stage-$1$ category reports, randomize independently of types and reports between $\widetilde M_m$ and $\bar M$ with probabilities $1-\eta_m$ and $\eta_m$. Temporarily disclose the selected branch before continuation. Both branches obey exact within-prior cost IC and truthful signal obedience; the first-stage report cannot depend on the branch. Thus maximum cross-prior gains in the mixture are bounded by the weighted average of the branchwise $D_{ss'}^i$. Restore the boundary rents
\(
K_{i,s}^{(m)}=(1-\eta_m)K_{i,s}+\eta_m\bar K_{i,s}.
\)
For every directed edge $(i,s,s')$, the restored cross-prior slack is at least
\(
(1-\eta_m)(-d_m)+\eta_m\rho=0.
\)
Consequently the restored mechanism is fully dynamically BIC.
Its truthful utility is the nonnegative normalized cost envelope
plus $K_{i,s}^{(m)}\ge0$, proving interim IR. Conditional on complete
category reports and experiment profiles, average out the branch
and other auxiliary pre-signal information by
Lemma~\ref{lem:auxiliary-averaging}; this preserves revenue and
does not increase deviation values.
By Lemma~\ref{lp:approx}, the discretized normalized branch loses at most $2n\bar C \delta_m$ in truthful expected revenue. Its unnormalized restored revenue is therefore at least \(\mathcal V(\widetilde M)-\sum_{i,s}\lambda_sK_{i,s}-2n\bar C \delta_m.\) The benchmark's restored revenue is $J_0$, proving \eqref{lp:repair-rev}.
For payment bounds: $|P_i^{\widetilde M}|\le B_1=b+2B_0$, the cell adjustment contributes at most $2\bar C \delta_m$, the normalized benchmark payment at most $b$, and $0\le K_{i,s}^{(m)}\le b+B_0$ by the hypothesis on the rent vector. Thus \(B=B_1+2\bar C+b+B_0=2b+3B_0+2\bar C\) bounds the intermediate repaired mechanism. Lemma~\ref{lem:finite-category-transfer-normalization} subsequently reduces this bound to $B_0$ without lowering revenue.
\end{proof}

\begin{proof}[Proof of Theorem~\ref{thm:finite-experiment-LP}]
The weighted-variable constraints \eqref{lp:feas}--\eqref{lp:obj} define a finite LP with payment bound $B_0$. For each actual cost cell, every fixed stage-$1$/continuation deviation has utility affine in the actual cost. The endpoint inequalities \eqref{lp:ic}--\eqref{lp:ir} are therefore sufficient and necessary for exact dynamic BIC and IR at every cost, including shared endpoints. Recovering primitive rules from the weighted variables on positive-probability experiment branches (and choosing arbitrary feasible rules otherwise) proves that $R_m^{\mathrm{LP}}(B_0)\le R^*$.

To prove the other inequality, take arbitrary $M\in\mathcal M^0$.
Lemma~\ref{lem:bounded-reduced-payments} gives
$M^0\in\mathcal M^0(B_0)$ with revenue no smaller than
$\operatorname{Rev}(M)$. Normalize its boundary rents as in the
proof of Proposition~\ref{prop:finite-prior-supremum}; since
$0\le K_{i,s}=U_i^{M^0}(s,1)\le b+B_0$, the resulting
$\widetilde M$ belongs to $\mathcal N^0(B_1)$. 
The pair is
feasible and
$\mathcal V(\widetilde M)-\sum_{i,s}\lambda_sK_{i,s}
=\operatorname{Rev}(M^0)$. 
Lemma~\ref{lp:approx} and lemma
~\ref{lp:mix} construct an exact-IC, interim-IR finite-category
mechanism for every cost type, with intermediate payment bound $B$
and revenue at least
\(
(1-\eta_m)[\operatorname{Rev}(M)-2n\bar C \delta_m]+\eta_mJ_0.
\)
Apply Lemma~\ref{lem:finite-category-transfer-normalization} to obtain a finite-category mechanism with payments bounded by $B_0$, no smaller revenue, and exact dynamic BIC and interim IR. 
It is feasible for $\mathrm{LP}_m(B_0)$, so taking the supremum over all $M$ gives
\(
R_m^{\mathrm{LP}}(B_0)\ge(1-\eta_m)(R^*-2n\bar C \delta_m)+\eta_mJ_0.
\)
Together with $R_m^{\mathrm{LP}}(B_0)\le R^*$ this yields
\[
0\le R^*-R_m^{\mathrm{LP}}(B_0)
\le (1-\eta_m)2n\bar C \delta_m+\eta_m(R^*-J_0)
\le2n\bar C \delta_m+\frac{3\bar C \delta_m}{\rho+3\bar C \delta_m}(R^*-J_0).
\]
Since $\rho>0$ is fixed, as $\delta_m\to0$ the optimal finite-LP revenue, $R_m^{\mathrm{LP}}(B_0)$, converges to \(R^*\) at a rate proportional to the maximal interval length \(\delta_m\).
\end{proof}

\subsubsection*{Joint finite-LP optimality conditions}
\begin{proof}[Proof of Theorem~\ref{thm:joint-finite-LP-rules}]
Write $\Sigma_i$ for the finite set of maps $\sigma_i:\Pi\times Y\to Y$. Each IC multiplier $\beta_{i,s,k,j;h_i',\sigma_i}$ corresponds to the slack
\(
A_i(s;(s,k),\mathrm{id})-t_jQ_i(s,k)
-A_i(s;h_i',\sigma_i)+t_jQ_i(h_i'),
\)
and $\eta_{i,s,k,j}$ corresponds to $A_i(s;(s,k),\mathrm{id})-t_jQ_i(s,k)$. All these multipliers are nonnegative. For $h_i=(s_i,k_i)$ set $w_i(h_i):=\lambda_{s_i}F_r(I_{k_i})$ and
\begin{align*}
\kappa_{i,h_i}&:=\sum_{j\in\{k_i-1,k_i\}}
\left(\eta_{i,s_i,k_i,j}
+\sum_{h_i'\in H_m,\,\sigma_i\in\Sigma_i}\!\beta_{i,s_i,k_i,j;h_i',\sigma_i}\right),\\
\ell^{\mathrm{tr}}_{i,h_i}&:=\sum_{j\in\{k_i-1,k_i\}}t_j
\left(\eta_{i,s_i,k_i,j}
+\sum_{h_i'\in H_m,\,\sigma_i\in\Sigma_i}\!\beta_{i,s_i,k_i,j;h_i',\sigma_i}\right),\\
\ell^{\mathrm{dev}}_{i,h_i}&:=\sum_{s\in\mathcal S}\sum_{k=1}^m
\sum_{j\in\{k-1,k\}}\sum_{\sigma_i\in\Sigma_i}
 t_j\beta_{i,s,k,j;h_i,\sigma_i}.
\end{align*}
For a deviation strategy $\sigma_i$, define the induced distributions over continuation reports by
\[m_s^{\pi_i,\sigma_i}(\widehat y_i)
:=\sum_{\{y_i:\sigma_i(\pi_i,y_i)=\widehat y_i\}}m_s^{\pi_i}(y_i)\quad \text{and}\quad n_s^{\pi_i,\sigma_i}(\widehat y_i)
:=\sum_{\{y_i:\sigma_i(\pi_i,y_i)=\widehat y_i\}}n_s^{\pi_i}(y_i).\]
For $h=(h_i,h_{-i})$ and $\widehat y\in Y^n$, abbreviate
\(
M_{-i}(h,\pi,\widehat y):=
 w_{-i}(h_{-i})\prod_{\ell\ne i}
 m_{s_\ell}^{\pi_\ell}(\widehat y_\ell).
\)
Expanding the linear forms \eqref{lp:Adef}--\eqref{lp:Qdef}, the IC and IR slacks \eqref{lp:ic}--\eqref{lp:ir}, and the revenue objective \eqref{lp:obj} gives the following coefficients of the weighted variables:
\begin{align}
 a_{h,\pi}
 &:=\sum_i w_{-i}(h_{-i})C(\pi_i)
 \bigl(\ell^{\mathrm{dev}}_{i,h_i}-\ell^{\mathrm{tr}}_{i,h_i}\bigr),\label{eq:joint-lp-a}\\
 c_{i,h,\pi,\widehat y}
 &:=M_{-i}(h,\pi,\widehat y)
 \left[\kappa_{i,h_i}n_{s_i}^{\pi_i}(\widehat y_i)
 -\sum_{s,k,j,\sigma_i}
 \beta_{i,s,k,j;h_i,\sigma_i}
 n_s^{\pi_i,\sigma_i}(\widehat y_i)\right],\label{eq:joint-lp-c}\\
 d_{i,h,\pi,\widehat y}
 &:=M_{-i}(h,\pi,\widehat y)
 \left[(w_i(h_i)-\kappa_{i,h_i})m_{s_i}^{\pi_i}(\widehat y_i)
 +\sum_{s,k,j,\sigma_i}
 \beta_{i,s,k,j;h_i,\sigma_i}
 m_s^{\pi_i,\sigma_i}(\widehat y_i)\right].\label{eq:joint-lp-d}
\end{align}
In the last two expressions, $\sum_{s,k,j,\sigma_i}$ ranges over $s\in\mathcal S$, $k=1,\ldots,m$, $j\in\{k-1,k\}$, and $\sigma_i\in\Sigma_i$. In particular, the term $w_i(h_i)$ in \eqref{eq:joint-lp-d} comes from the truthful-revenue objective, the terms with $\kappa_{i,h_i}$ from truthful IC and IR payoffs, and the remaining terms from deviations into report $h_i$. The cost coefficients \eqref{eq:joint-lp-a} come only from the cost terms in the incentive and participation constraints.

For fixed $(\beta,\eta)$, maximize this Lagrangian subject only to the primitive weighted feasibility restrictions \eqref{lp:feas}. If $\gamma_{h,\pi}>0$, write $\xi_{i,h,\pi,\widehat y}=\gamma_{h,\pi}p_i(\widehat y\mid h,\pi)$ and $\zeta_{i,h,\pi,\widehat y}=\gamma_{h,\pi}P_i(\widehat y\mid h,\pi)$. The allocation simplex yields
\[
\max_{p_i\ge0,\ \sum_i p_i\le1}
 \sum_i c_{i,h,\pi,\widehat y}p_i
 =\max\{0,\max_i c_{i,h,\pi,\widehat y}\},
\]
and independent payment bounds $|P_i|\le B_0$ yield
\[
\max_{|P_i|\le B_0}\sum_i d_{i,h,\pi,\widehat y}P_i
 =B_0\sum_i|d_{i,h,\pi,\widehat y}|.
\]
These maxima hold for every report profile, experiment profile, and continuation report; for $\gamma_{h,\pi}=0$ all corresponding weighted variables vanish. Hence, after optimizing continuation rules, the Lagrangian equals
$\sum_{h,\pi}\gamma_{h,\pi}\mathcal J_h(\pi;\beta,\eta)$.
Maximizing each probability vector $(\gamma_{h,\pi})_{\pi\in\Pi^n}$ over its simplex gives
\[
\sup_{(\gamma,\xi,\zeta)\text{ satisfying }\eqref{lp:feas}}
\mathcal L(\gamma,\xi,\zeta;\beta,\eta)
=\sum_h\max_{\pi\in\Pi^n}\mathcal J_h(\pi;\beta,\eta).
\]
The finite LP is feasible (assign the zero-cost uninformative experiment, allocate nothing, and charge zero) and bounded by \eqref{lp:feas}. Finite-dimensional LP strong duality therefore gives \eqref{eq:joint-lp-value-dual} and attainment of the dual minimum.

At a primal-optimal mechanism and dual-optimal multipliers, the primal and dual values coincide. Since the IC/IR slacks and their multipliers are nonnegative, the equality forces complementary slackness and forces the mechanism to maximize the Lagrangian over \eqref{lp:feas}. By the pointwise maximizations above, this is equivalent to conditions (i)--(iii), with (iv) giving complementary slackness. Conversely, for a feasible mechanism satisfying (i)--(iv) with nonnegative multipliers, its Lagrangian equals its revenue and attains the Lagrangian supremum. The weak-duality upper bound then equals that revenue, proving primal and dual optimality.
\end{proof}

\section*{Supplementary Appendix}
The following calculations provide revenue bounds, mechanism
constructions, and complete incentive and participation checks
for the two-prior illustration in
Section~\ref{par:two-prior-example}.
The proofs of the general results appear in the preceding appendix.

\appendix
\renewcommand{\thesection}{S.\arabic{section}}
\renewcommand{\theHsection}{supp.\arabic{section}}
\renewcommand{\theHequation}{supp.\arabic{section}.\arabic{equation}}
\numberwithin{equation}{section}

\section{A two-prior comparison of continuation rules}\label{supp:two-prior-comparison}

For the two-prior illustration in Section~\ref{par:two-prior-example}, let $L=1/5$, $H=2/5$, with equal prior probabilities and independent uniform costs. The uninformative $\pi^0$ always gives signal $-$ at zero cost; $\pi^1$ has $\pi^1(+\mid1)=6/7$, $\pi^1(+\mid0)=1/7$ and $C(\pi^1)=1/245$. Its signal probabilities and posterior values are
\[
\begin{array}{c|ccc}
 s & m_s^{\pi^1}(+) & t(s,\pi^1,+) & t(s,\pi^1,-)\\ \hline
 L&2/7&3/5&1/25\\
 H&3/7&4/5&1/10
\end{array}
\]

\emph{A feasible non-VCG mechanism.}
Give both bidders $\pi^1$, independently of their reports, and use
a second-price auction with reserve $P_{\mathrm{res}}=3/5$ on the
posterior bids $t(\widehat s_i,\pi^1,\widehat y_i)$ computed from
their reported priors and signal labels. Subtract $1/245$ from
each bidder's continuation payment, a subsidy that covers
$r_iC(\pi^1)$ for all $r_i\in[0,1]$. Truthful stage-$1$ prior
reports and truthful signal reports jointly reproduce truthful
posterior bids. Since bidding one's true posterior is optimal in a
second-price auction with reserve against the opponents' truthful
bids, no joint stage-$1$/continuation misreport improves expected
gross auction payoff. Experiment assignments and the subsidy ignore
cost reports, establishing dynamic BIC and interim IR in
$\mathcal M^0$.

The unconditional
probability of a negative signal is $9/14$. A payment above $P_{\mathrm{res}}$ occurs
only when both bidders have prior $H$ and positive signals, with
probability $(3/14)^2$; then the payment is $4/5$. Hence
\[
R^{\mathrm{res}}=
\frac35\left[1-\left(\frac9{14}\right)^2\right]
+\left(\frac45-\frac35\right)\left(\frac3{14}\right)^2
-\frac2{245}
=\frac{173}{490}.
\]

Averaging over the two equally likely prior types, the
unconditional distribution of posterior values assigns
probabilities $5/14,4/14,2/14,3/14$ to
$1/25,1/10,3/5,4/5$, respectively. The first two
(ironed) Myerson virtual values are negative and the
last two positive and increasing. Hence the reserve
auction is Myerson-optimal for this unconditional
distribution; its rules do not depend on prior reports,
and it remains dynamically BIC.

\emph{Upper bound for the posterior-bid VCG benchmark.}
For this separate benchmark, the continuation report is an
unrestricted posterior-value bid, and the continuation rule
is the VCG second-price rule. This is the standard VCG
formulation and is used here only as a comparison, not as
the primitive continuation-report space $Y$.
For $s,t\in\{L,H\}$ and a technology profile $\pi$, the expected efficient gross surplus is $G((s,t),\pi)$, as defined in \eqref{eq:gross-surplus}. Define
\(
A_1(s,t;\pi):=E[(T_1-T_2)_+\mid s,t,\pi],
\)
with $A_2$ defined symmetrically, and let
\(d_i(t;\pi):=A_i(H,t;\pi)-A_i(L,t;\pi),\) where $t$ is bidder $i$'s
opponent's prior. Under any report-dependent VCG continuation,
fix bidder $i$'s cost report $r$ and let a true $H$ type deviate
by reporting $(L,r)$ and then bidding her true posterior in the
continuation auction. Relative to the same stage-$1$ report by
a true $L$ type, the payment and information-cost terms coincide.
By stage-$1$
IC and interim IR for $L$,
\[
U_i(H,r)\ge U_i(L,r)
+E\bigl[d_i(s_{-i};\pi(L,r,\theta_{-i}))\bigr]
\ge E\bigl[d_i(s_{-i};\pi(L,r,\theta_{-i}))\bigr],
\]
where for randomized assignments the expression is averaged over their
realizations. Summing over bidders and true types, dropping nonnegative
information costs, and using revenue as expected gross surplus minus
bidder utility shows that VCG revenue is at most the expectation, over
prior and cost reports, of
\(
G((s_1,s_2),\pi)
-\sum_{i:\,s_i=L}d_i(s_{-i};\pi).
\)
For each reported prior profile this expression is linear in the
assignment lottery, so it is bounded by its largest value across the
four pairs $\pi\in\{\pi^0,\pi^1\}^2$. Direct substitution of the
above posterior distributions gives the following values (the columns
specify which bidders receive $\pi^1$):
\[
\begin{array}{c|cccc|c}
(s_1,s_2)&(0,0)&(0,1)&(1,0)&(1,1)&\max\\ \hline
(L,L)&-1/5&1/35&1/35&-1/49&1/35\\
(L,H)&2/5&12/35&12/35&96/245&2/5\\
(H,L)&2/5&12/35&12/35&96/245&2/5\\
(H,H)&2/5&4/7&4/7&4/7&4/7
\end{array}
\]
Each prior profile has probability $1/4$, implying
\[
\operatorname{Rev}^{\mathrm{VCG}}
\le\frac14\left(\frac1{35}+\frac25+\frac25+\frac47\right)
=\frac7{20}<R^{\mathrm{res}}.
\]
The bound permits arbitrary report-dependent correlated experiments
and stage-$1$ transfers in this posterior-bid VCG comparator,
subject to dynamic BIC and interim IR. It is not asserted as a
bound for all efficient allocation rules in $\mathcal M^0$.

\emph{Optimal continuation with fixed informative learning.}
Now require $\pi^1$ to be assigned to both bidders for every stage-$1$
report, while allowing arbitrary admissible continuation allocation
and signed payments. The optimal revenue in this
fixed-information class is
\begin{equation}\label{eq:fixed-information-example-optimum}
R_{\mathrm{fixed}}^*=\frac{19}{49}.
\end{equation}
Put $\varepsilon:=1/245$ and, for $s\in\{L,H\}$, write
\(
m_s(y):=\Pr(Y_i=y\mid s_i=s)\) and \(v_{s,y}:=t(s,\pi^1,y).
\)
For a fixed bidder and a report $(s,r)$, let $Q_{s,y}(r)$ and
$P_{s,y}(r)$ be her expected allocation probability and \emph{total}
continuation payment conditional on observing signal $y$, with
opponents truthful. Independence gives the same conditional quantities
after prior misreports. The bound permits report- and signal-dependent
payments.

Since the experiment is fixed, changing the true cost scale changes
payoffs by $-r\varepsilon$ regardless of the report. Within-prior
cost-report IC thus makes the \emph{gross} truthful interim utility
\[
g_s=\sum_{y\in\{-,+\}}m_s(y)
   [v_{s,y}Q_{s,y}(r)-P_{s,y}(r)]
\]
independent of $r$. Interim IR for the type with $r=1$ implies
$g_L\geq\varepsilon$. 

Fix a cost report $r$ and abbreviate
$Q_{L,y}(r)$ and $P_{L,y}(r)$ by $Q_y$ and $P_y$. A true $H$ bidder can
report $(L,r)$ and, following her realized signal, send the continuation
message used by an $L$ bidder with the same signal. Dynamic IC therefore
gives
\begin{align*}
g_H-g_L
&\geq \tfrac1{35}Q_-+\tfrac6{35}Q_+
        -\tfrac17(P_+-P_-)\\
&\geq \tfrac4{35}Q_-+\tfrac3{35}Q_+.
\end{align*}
The second inequality follows from continuation obedience of an
$L,+$ bidder:
\(
P_+-P_-\leq \frac35(Q_+-Q_-).
\)
These deviations are admissible in the primitive signal-label class.
For the following bound, let $Q_{s,y}$ denote the average of
$Q^i_{s,y}(r)$ over bidders $i=1,2$ and uniformly distributed cost
reports $r$. Summing the preceding inequalities over bidders, revenue
accounting and $g_L\geq\varepsilon$ give
\[
\operatorname{Rev}(M)\leq
-2\varepsilon
+\frac3{35}(Q_{L,+}-Q_{L,-})
+\frac2{35}Q_{H,-}+\frac{12}{35}Q_{H,+}.
\]
Equivalently, if $\phi_{s,y}$ are the following rent-adjusted values,
\[
\phi_{L,-}=-\frac3{25},\;
\phi_{H,-}=\frac1{10},\;
\phi_{L,+}=\frac3{10},\;
\phi_{H,+}=\frac45,
\]
then the right-hand side is
$-2\varepsilon+E[\sum_i\phi_{s_i,Y_i}p_i]$.
Feasibility implies that the latter expectation cannot exceed
$E[\max\{0,\phi_{s_1,Y_1},\phi_{s_2,Y_2}\}]$. The four posterior
labels \((L,-),(H,-),(L,+),(H,+)\) have unconditional probabilities
$(5,4,2,3)/14$, respectively. Direct evaluation yields
\[
\operatorname{Rev}(M)
\leq -\frac2{245}
+E[\max\{0,\phi_{s_1,Y_1},\phi_{s_2,Y_2}\}]
=-\frac2{245}+\frac{97}{245}
=\frac{19}{49}.
\]

The same inequality also establishes a strict gap for efficient
continuation \emph{within the primitive signal-label class} when
both experiments are fixed at $\pi^1$. With probability
$[(1/2)(5/7)]^2=25/196$ both realized prior--signal labels
are $(L,-)$. On this event, an efficient continuation must
allocate the object to a bidder, because posterior value
$1/25$ is strictly positive. Both rent-adjusted values are
$\phi_{L,-}=-3/25$, so the allocation contributes $-3/25$
instead of the maximum feasible rent-adjusted contribution $0$.
At every other profile the allocation contribution remains bounded
above by the positive-part maximum. Thus every dynamically BIC,
interim-IR mechanism with fixed informative experiments and
efficient continuation obeys
\[
\operatorname{Rev}^{\mathrm{eff}}_{\mathrm{fixed}}
\le\frac{19}{49}-\frac{25}{196}\frac3{25}
=\frac{73}{196}<\frac{19}{49}.
\]
This argument does not use the posterior-bid VCG comparator.

To attain the bound, use the primitive triple $(\Lambda,p,P)$.
At stage $2$, allocate to the largest positive rent-adjusted
value $\phi_{s,y}$, breaking ties uniformly, and use the following
\emph{own-message} continuation payments, whether or not the bidder
receives the object. Both signal labels are available under each
reported prior, so no off-grid continuation reports are needed:
\[
\begin{array}{c|c|c}
(s,y)&Q_{s,y}&P_{s,y}\\\hline
(L,-)&0&-1/245\\
(H,-)&1/2&3/20\\
(L,+)&5/7&104/245\\
(H,+)&25/28&3/7
\end{array}
\]
The implied gross truthful utilities are
$g_L=1/245$ and $g_H=16/245$.
For every cost type $r\in[0,1]$, the best expected continuation payoff
after reporting the other prior (allowing the continuation signal message
to be chosen optimally) is $-1/20$ for a true $L$ bidder
reporting $H$, and $16/245$ for a true $H$ bidder reporting $L$.
Within either reported prior, truthful signal reporting is optimal:
the $L$ positive-signal type is indifferent between the two signal reports,
whereas the $H$ positive-signal type strictly prefers reporting $+$;
both negative-signal types strictly prefer their truthful reports.
Independence from cost reports yields cost-report IC; $g_L\ge\varepsilon$ and $g_H>\varepsilon$ ensure interim IR. The mechanism is dynamically BIC and attains the upper bound. The allocation rule
withholds the object only if both realized labels are $(L,-)$; in
particular, it withholds when two low-prior bidders receive negative
signals, but sells when both high-prior bidders receive negative
signals. This proves \eqref{eq:fixed-information-example-optimum}.

\emph{Optimality among cost-independent experiment assignments.}
We strengthen \eqref{eq:fixed-information-example-optimum}: the same
revenue is optimal even if the experiment-assignment lottery may
vary with reported priors, as long as it is independent of reported
information-cost types. Continuation allocation and payments may
still depend on both cost reports, and assignments may be correlated.
Formally, the restriction on the assignment kernel is
\(
\Lambda(\pi\mid \hat s_1,\hat r_1,\hat s_2,\hat r_2)
 =\Lambda(\pi\mid \hat s_1,\hat s_2).
\)
All other admissible continuation rules remain unrestricted.
We show that every dynamically BIC and interim-IR mechanism in this class
has revenue at most $19/49$.

Fix a bidder $i$ and let $q_i(s)$ be her expected base experiment cost
when reporting prior $s$, with the opponent truthful. The assignment
restriction makes $q_i(s)$ independent of her reported cost. Let
$g_i(s,u)$ be her expected gross payoff (allocated value minus total
continuation payments) from reporting $(s,u)$ and
following the prescribed continuation under true prior $s$.
Within-prior cost-report IC implies
\(
 g_i(s,u)=g_i(s)
\) for every \(u\in[0,1]\),
since the learning-cost term $r_iq_i(s)$ is the same for all own
cost reports at the fixed prior $s$. Interim IR for $(L,1)$ gives
$g_i(L)\ge q_i(L)$.

Consider a true $H$ bidder with $r_i=0$ who instead reports $(L,u)$
and then imitates the continuation messages of a truthful $L$ bidder
with her realized signal. The distribution of the assigned experiment
and opponents' variables is the same under this
report for true prior $L$ and true prior $H$. If her assigned experiment
is $\pi^0$, the resulting gross-payoff difference relative to a true
$L$ bidder making the same report is $\frac15 Q_0$, where $Q_0$ is
her conditional allocation probability. If it is $\pi^1$, write
$Q_-,Q_+$ and $P_-,P_+$ for the expected allocations and total
payments from the respective low-prior truthful continuation messages,
conditional on the bidder's own assigned experiment.

By the maintained independence structure, the conditional laws of
opponents' variables are the same for the two own signals.
Continuation obedience for a true $(L,+)$ bidder yields
\(
 P_+-P_-\le \frac35(Q_+-Q_-).
\)
The corresponding difference in gross payoffs between true prior $H$
and true prior $L$, after both report $L$, satisfies
\begin{align*}
 D_i
 &=\frac1{35}Q_-+\frac6{35}Q_+-\frac17(P_+-P_-)\\
 &\ge \frac4{35}Q_-+\frac3{35}Q_+.
\end{align*}
These inequalities hold conditional on her assigned experiment,
and hence after integrating over the opponent and the
assigned experiment. For each reported cost $u$, let $d_i(u)$ denote
the resulting lower bound on this cross-prior gross-payoff difference,
using coefficient $1/5$ for the uninformed experiment and coefficients
$4/35,3/35$ for negative and positive informative signals.

Stage-$1$ IC of the $(H,0)$ bidder implies
\(
 g_i(H)\ge g_i(L)+d_i(u)
\) for all \(u\).
Average this inequality over $u\sim U[0,1]$, writing
$\bar d_i=\int_0^1d_i(u)\,du$. Together with low-prior IR,
that
\(
 g_i(L)+g_i(H)\ge 2q_i(L)+\bar d_i
\) holds.
Since priors are equally likely, expected revenue therefore satisfies
\begin{equation}\label{eq:CI-interim-bound}
 \operatorname{Rev}(M)
 \le \mathbb E\Big[\sum_i V_i p_i\Big]
    -\sum_i q_i(L)-\frac12\sum_i\bar d_i.
\end{equation}

For an assignment $\pi^e$, $e\in\{0,1\}$, put
\[m_s^e(y)=\Pr(Y=y\mid s,\pi^e)\;\text{and}\;
n_s^e(y)=\mathbb E[V_i\mathbf1_{\{Y_i=y\}}\mid s,\pi^e].\]
The adjustment in \eqref{eq:CI-interim-bound} can be incorporated into
a \emph{rent-adjusted value} $\varphi_{s,e}(y)$, defined on positive-probability
signal histories by
\[
\begin{array}{c|ccc}
 &\pi^0,-&\pi^1,-&\pi^1,+\\ \hline
 L&0&-3/25&3/10\\
 H&2/5&1/10&4/5
\end{array}.
\]
For the informative experiment, these are the same rent-adjusted
values as in the fixed-information calculation:
$\varphi_{s,1}(y)=\phi_{s,y}$ for $s\in\{L,H\}$ and
$y\in\{-,+\}$.
For $s=L$, the adjusted signal-value numerators are
\(
n_L^0(-)-\tfrac15=0,\) 
\(n_L^1(-)-\tfrac4{35}=-\tfrac3{35},\) and
\(n_L^1(+)-\tfrac3{35}=\tfrac3{35}.
\)
Dividing by the corresponding signal probabilities yields the displayed
values. For $s=H$, the rent-adjusted values coincide with the ordinary posterior values. The term $\sum_iq_i(L)$ in
\eqref{eq:CI-interim-bound} contributes a cost penalty
$2C(\pi_i)$ for each low-prior bidder in a given prior profile,
because that profile has unconditional probability $1/4$ while
$q_i(L)$ is conditional on $s_i=L$, with probability $1/2$.

Let $e=(e_1,e_2)$ denote the assigned experiment profile and define
\[
 W^{\mathrm{CI}}(s_1,s_2;e_1,e_2)
 :=\mathbb E\!\left[
 \max\{0,\varphi_{s_1,e_1}(Y_1),
          \varphi_{s_2,e_2}(Y_2)\}
 \middle|s_1,s_2,e_1,e_2\right]
 -\frac2{245}\sum_{i:s_i=L}e_i.
\]
Averaging allocation rules over cost reports
conditional on the complete prior, experiment and signal profiles
preserves feasibility, so \eqref{eq:CI-interim-bound} and
$\sum_ip_i\le1$ imply
\[
 \operatorname{Rev}(M)
 \le\frac14\sum_{(s_1,s_2)\in\{L,H\}^2}
       \max_{(e_1,e_2)\in\{0,1\}^2}W^{\mathrm{CI}}(s_1,s_2;e_1,e_2).
\]
The exact contributions $\frac14W^{\mathrm{CI}}$ are
\[
\begin{array}{c|cccc}
(s_1,s_2)&(0,0)&(0,1)&(1,0)&(1,1)\\\hline
(L,L)&0&19/980&19/980&8/245\\
(L,H)&1/10&1/10&24/245&26/245\\
(H,L)&1/10&24/245&1/10&26/245\\
(H,H)&1/10&1/7&1/7&1/7
\end{array}.
\]
Summing the row maxima yields
\(
 \operatorname{Rev}(M)
 \le \frac8{245}
      +2\frac{26}{245}+\frac17
 =\frac{19}{49}.
\)
The fixed-informative mechanism constructed above attains this value.
Thus $R_{\mathrm{CI}}^*=R_{\mathrm{fixed}}^*$.
The cost-contingent mechanism $\hat M$ constructed
below establishes
\(
\operatorname{Rev}(\hat M)-R_{\mathrm{CI}}^*
=\frac{3}{7840}>0.
\)

\emph{A strict revenue gain from cost-contingent experimentation.}
We establish the strict improvement in \eqref{eq:two-prior-endogenous-revenue} by constructing
an explicit mechanism. Interpret each cost report through the two categories
$0$ for $r\leq 1/2$ and $1$ for $r>1/2$; write $L_0,L_1,H_0,H_1$
for the resulting prior--cost reports. The mechanism uses the
primitive $(\Lambda,p,P)$ representation. Assign $\pi^1$ to both
bidders unless both report
prior $H$ and at least one reports cost category $1$. For reports
$(H_0,H_1)$ or $(H_1,H_0)$, inform only the bidder reporting category
$0$. For $(H_1,H_1)$, inform one bidder uniformly at random.
If both bidders are informed, allocate the object to the largest strictly
positive score $\phi_{s,y}$, breaking ties uniformly, where
\(
(\phi_{L,-},\phi_{H,-},\phi_{L,+},\phi_{H,+})
=\left(-\frac3{25},\frac1{10},\frac3{10},\frac45\right).
\)
If exactly one bidder is informed, allocate to her when her reported
signal is $+$ and to the uninformed bidder otherwise.
The continuation payment depends only
on the bidder's own reported category, assigned experiment, and signal
label; it is given by the following table (payments are made to the seller):
\[
\begin{array}{c|c|rr}
\text{Reported category} & \text{Experiment} & \hat y=- & \hat y=+\\ \hline
L_0\text{ or }L_1 &\pi^1&-1/245&104/245\\
H_0&\pi^1&1/7&33/70\\
H_1&\pi^1&2/7&4/7\\
H_1&\pi^0&1/490&1/490
\end{array}
\]
An uninformed bidder's allocation is also independent of her signal
report, including the off-support label $+$.

For reference, the conditional allocation probabilities induced by this
assignment and ranking rule, averaged over truthful opponents and
conditional on the bidder receiving the indicated experiment, are
\[
\begin{array}{c|c|cc}
\text{Reported category}&\text{Experiment}&Q_{c,-}&Q_{c,+}\\\hline
L_0\text{ or }L_1&\pi^1&0&5/7\\
H_0&\pi^1&3/7&53/56\\
H_1&\pi^1&4/7&1\\
H_1&\pi^0&4/7&4/7
\end{array}.
\]
The last row is independent of the signal report because $\pi^0$ is
uninformative. The probability of assignment to $\pi^1$ is $1$ for
$L_0,L_1,H_0$ and $5/8$ for $H_1$.

For true prior $s$ and a reported category $c$, let $A_{s,c}$ denote the
maximum expected continuation gross payoff before the cost $rC(\pi_i)$,
optimizing over all continuation signal-report strategies conditional on
own assigned experiment and realized signal. Let $q_c$ be the expected
base learning cost. Conditional on a fixed own category report, average
over the opponent's four equally likely truthful categories, the specified
experiment lottery, and both signals. Comparing the two continuation
labels at every own experiment--signal history yields the exact values
\[
\begin{array}{c|rrrr}
 & L_0&L_1&H_0&H_1\\ \hline
A_{L,c}&1/245&1/245&-2/35&-51/784\\
A_{H,c}&16/245&16/245&16/245&253/3920\\
q_c&1/245&1/245&1/245&1/392
\end{array}
\]
Truthful signal reporting attains $A_{s,c}$ whenever the reported prior
equals $s$; thus continuation obedience holds. For a true $L$ bidder,
truthful interim utility is
\(
U_L(r)=(1-r)/245,
\)
which is nonnegative and dominates $A_{L,c}-rq_c$ for all four reports.
For a true $H$ bidder, the best available interim payoff is
\(
U_H(r)=\max\left\{
\frac{16}{245}-\frac r{245},\quad
\frac{253}{3920}-\frac r{392}\right\}.
\)
The two affine terms are equal at $r=1/2$: the first is achieved by
truthful report $H_0$ when $r\leq1/2$, and the second by truthful
report $H_1$ when $r>1/2$. The other two category reports yield the
first affine term. Hence no false-prior, false-cost-category, or
continuation deviation is profitable. Also $U_H(r)>0$ throughout
$[0,1]$, establishing interim IR. These comparisons cover every
continuous cost type, not only the category endpoints.

The expected payment from either $L$ category is $29/245$, from
$H_0$ it is $139/490$, and from $H_1$ it is $1003/3920$.
Each category occurs with probability $1/4$ for each bidder, so expected
seller revenue is
\[
\operatorname{Rev}(\hat M)
=\frac12\left(
2\cdot\frac{29}{245}
+\frac{139}{490}
+\frac{1003}{3920}\right)
=\frac{3043}{7840}
=\frac{19}{49}+\frac3{7840}.
\]
This verifies \eqref{eq:two-prior-endogenous-revenue}.

\end{document}